\documentclass[aps,prl,superscriptaddress,twocolumn,longbibliography]{revtex4-1}%,

\usepackage{amsfonts}
\usepackage{subfigure}
\usepackage{amsmath}
\usepackage{txfonts}
\usepackage{amssymb}
\usepackage{amsbsy} %for sigma in bold face
\usepackage{epsfig}
\usepackage{graphicx}

\def\be{\begin{equation}} \def\ee{\end{equation}}
\def\bea{\begin{eqnarray}} \def\eea{\end{eqnarray}}

\def\nn{\nonumber}

\def\bk{{\bf k}}

\def\bB{{\bf B}}

\def\be{{\bf e}}
\def\bd{{\bf d}}
\def\bA{{\bf A}}

\def\bn{{\bf n}}

\def\rw{\rightarrow}

\begin{document}

\title{Nodal-link semimetals}

\author{Zhongbo Yan}
\affiliation{ Institute for
Advanced Study, Tsinghua University, Beijing, 100084,  China}

\author{Ren Bi}
\affiliation{ Institute for
Advanced Study, Tsinghua University, Beijing, 100084,  China}

\author{Huitao Shen}
\affiliation{Department of physics, Massachusetts Institute of Technology, Cambridge, MA 02139, USA}
\affiliation{ Institute for
Advanced Study, Tsinghua University, Beijing, 100084,  China}

\author{Ling Lu}
\affiliation{ Institute of Physics, Chinese Academy of Sciences/Beijing National Laboratory for Condensed Matter Physics, Beijing 100190, China }

\author{Shou-Cheng Zhang}

\affiliation{
Department of Physics, Stanford University, CA 94305, USA}

\author{Zhong Wang}
\altaffiliation{  wangzhongemail@tsinghua.edu.cn} \affiliation{ Institute for
Advanced Study, Tsinghua University, Beijing, 100084,  China}

%\affiliation{Collaborative Innovation Center of Quantum Matter, Beijing, 100871, China }

%\date{\today}

\begin{abstract}

In topological semimetals, the valence band and conduction band meet at zero-dimensional nodal points or one-dimensional nodal rings, which are protected by band topology and symmetries. In this Rapid Communication,
we introduce ``nodal-link semimetals'', which host linked nodal rings in the Brillouin zone. We put forward a general recipe  based on the Hopf map for constructing models of nodal-link semimetal.  The consequences of nodal ring linking in the Landau levels and Floquet properties are investigated.

\end{abstract}

\maketitle

Topological phases of matter have been among the most active research subjects in condensed matter physics. They can be broadly classified as two major classes. The first class of phases, including topological insulators and superconductors\cite{hasan2010,qi2011,Chiu2015RMP,Bansil2016, bernevig2013topological,shen2013topological}, and other symmetry protected topological phases\cite{chen2013symmetry}, have gapped bulks with nontrivial topological structures characterized by topological invariants\cite{thouless1982,niu1985,kane2005b,fu2007b,moore2007,qi2008,schnyder2008,wang2012a}, dictating the existence of robust gapless modes on the boundary.

More recently, the second major class of topological materials, known as topological semimetals, have attracted widespread attentions. In the noninteracting limit, they are characterized by topologically robust $\bk$-space band-touching manifolds, which can be zero-dimensional (0D) nodal points or one-dimensional (1D) nodal rings (or nodal lines). The bulk Dirac\cite{liu2014discovery,neupane2014,Borisenko2014, xu2015observation,young2012dirac,
wang2012dirac,wang2013three,Zhang2015detection,
Chen2015Magnetoinfrared,liu2016zeeman} and Weyl points\cite{wan2011,volovik2003,yang2011,
burkov2011,weng2015,Huang2015TaAs,Xu2015weyl,lv2015,
YangLexian,Huang2015,Xu2015NbAs,Shekhar,lu2013weyl,lu2015,soluyanov2015type} are responsible for novel phenomena related to chiral anomaly\cite{son2012,liu2012,aji2011,zyuzin2012,wang2013a,Hosur2013,Hosur-anomaly,
Kim-chiral-anomaly,
Parameswaran-anomaly,Li2015ZrTe5,Bi2015}. Moreover,  bulk Weyl points entail surface Fermi arcs, while nodal rings\cite{Burkov2011nodal,Carter2012,
Phillips2014tunable,
Zeng2015nodal,chen2015topological,Chiu2014,Weng2015nodal,Mullen2015,Yu2015,Kim2015,Bian2015nodal,
Xie2015ring,Rhim2015Landau,Chen2015spin,
Fang2015nodal,Bian2015TlTaSe,yu2017topological,Rhim2016,Yan2016nodal,Lim2017nodal,fang2017} imply flat surface band (drumhead states) that may trigger interesting correlation effects\cite{Liu2017nodal}. Nodal rings have been predicted (e.g., Cu$_3$PdN\cite{Yu2015,Kim2015},Hg$_3$As$_2$\cite{Lu2016Node}, Ca$_3$P$_2$\cite{Xie2015ring,Chan2015Ca3P2}, 3D carbon networks\cite{Weng2015nodal}, ${\mathrm{CaP}}_{3}$\cite{Xu2017nodal}, Alkali Earth Metals\cite{Li2016alkai,Hirayama2017nodal}) and experimentally studied in quite a few materials (e.g., PbTaSe$_2$\cite{Bian2015nodal}, ZrSiTe\cite{hu2016ZrSiTe}, ZrSiS\cite{schoop2015dirac,Singha2016,Neupane2016,Wang2016evidence,Chen2017Dirac}). Notably, nodal rings can be driven to Floquet Weyl points by circularly polarized light\cite{yan2016tunable,Chan2016type,Narayan2016nodal,Zhang2016floquet,Taguchi2016nodal}, accordingly, the drumhead surface states become Fermi arcs.

\begin{figure*}[t!]
\subfigure{\includegraphics[width=5.4cm, height=4cm]{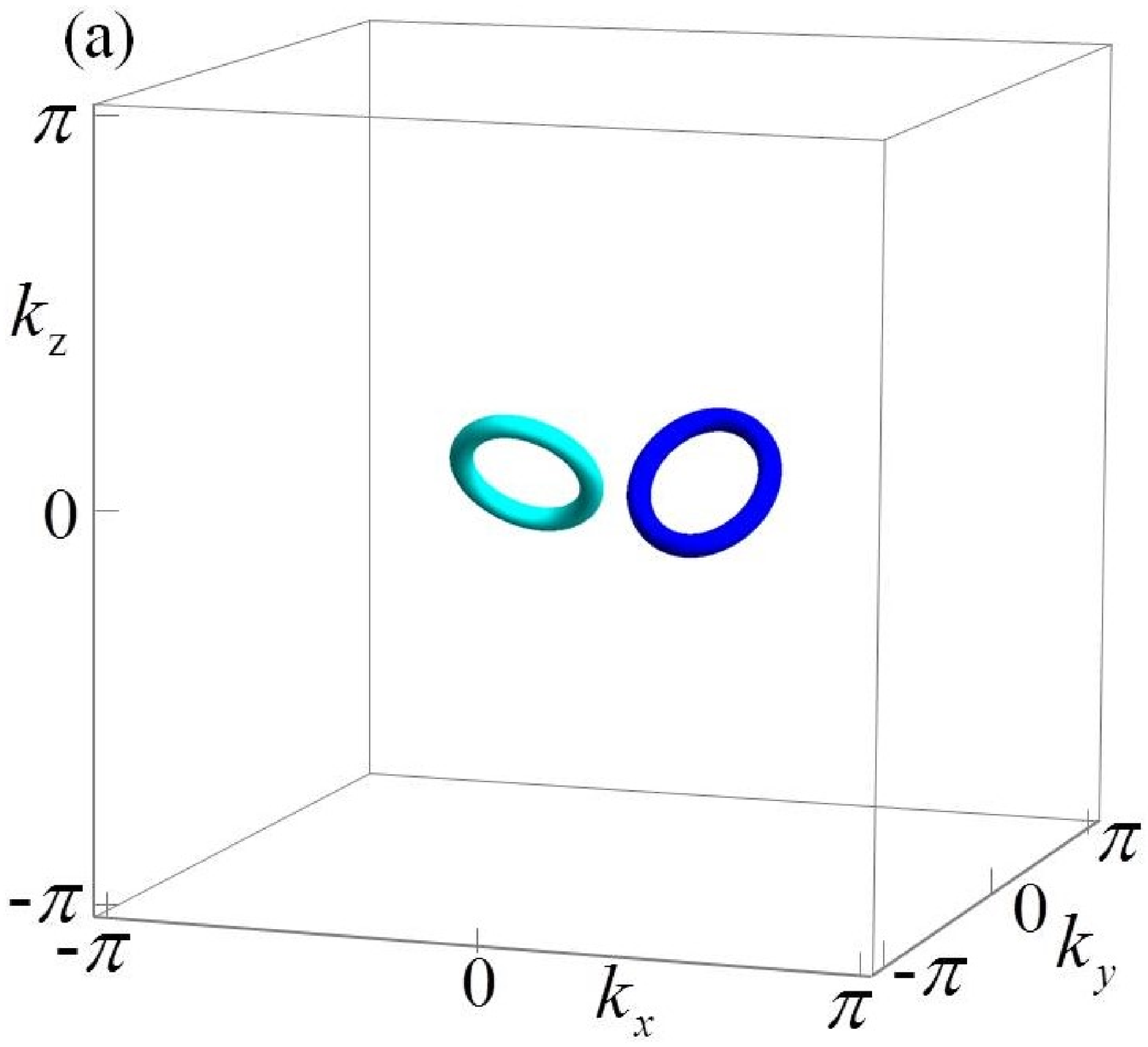}}
\subfigure{\includegraphics[width=5.4cm, height=4cm]{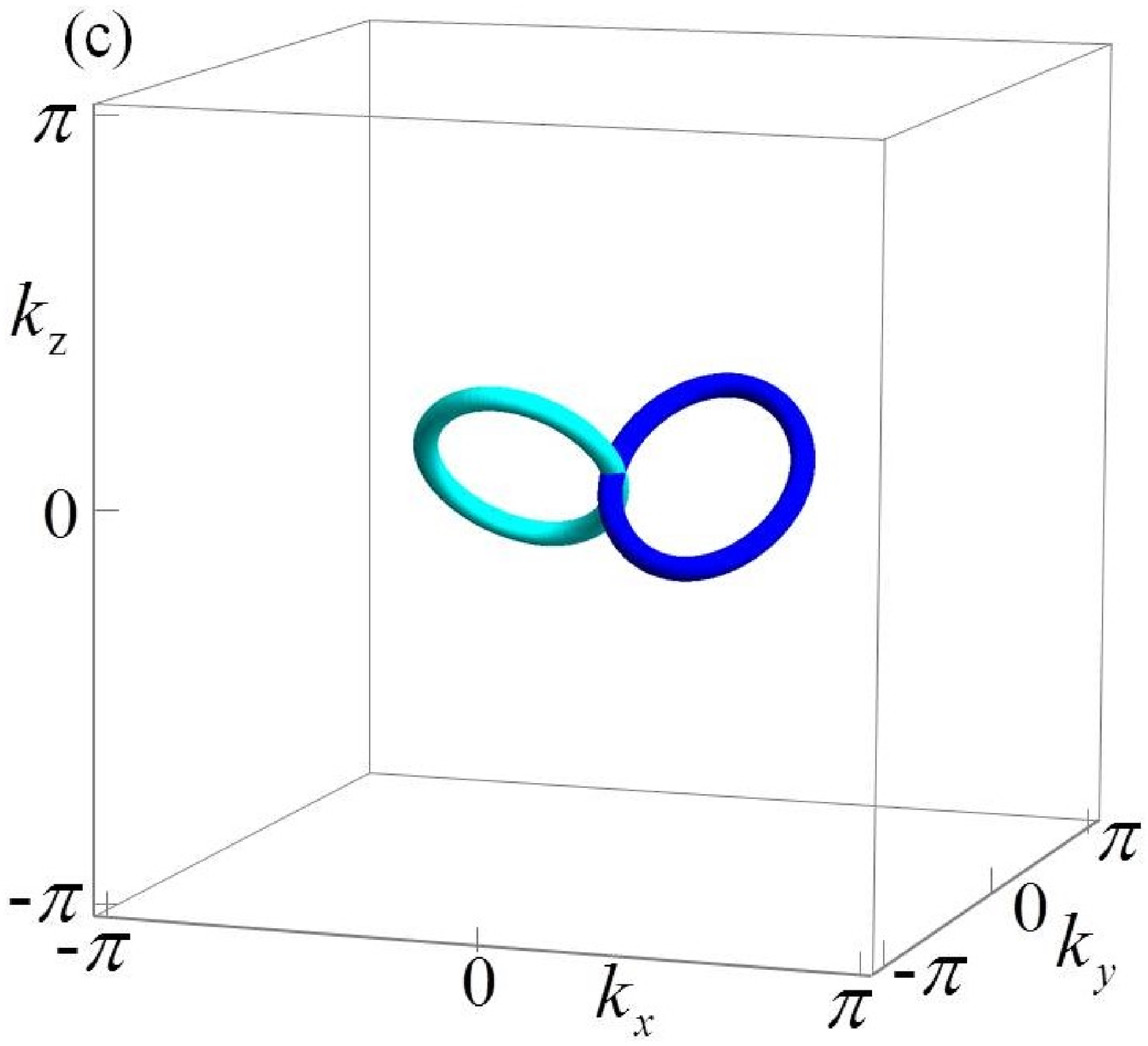}}
\subfigure{\includegraphics[width=5.4cm, height=4cm]{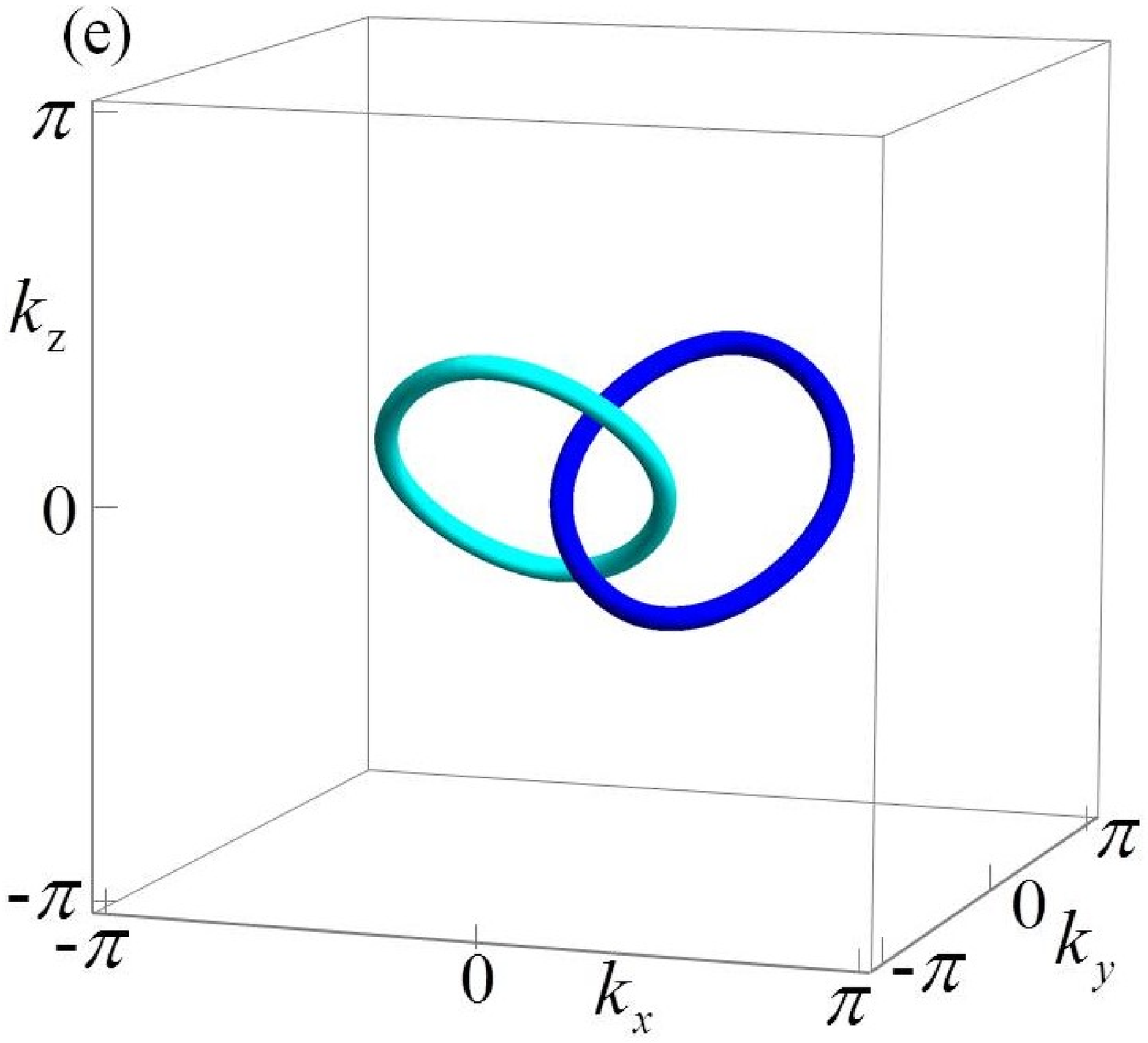}}
\subfigure{\includegraphics[width=5.4cm, height=4cm]{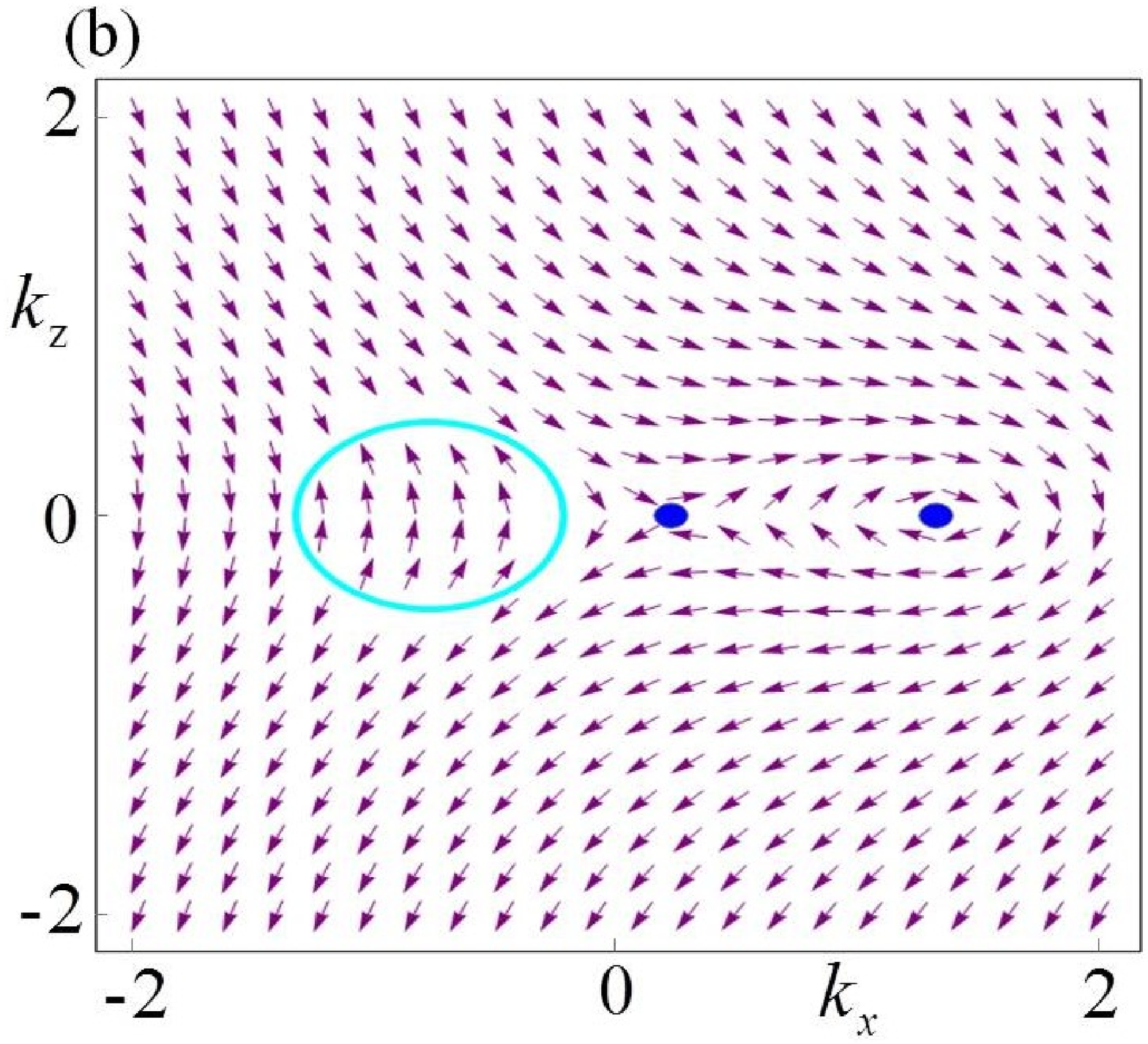}}
\subfigure{\includegraphics[width=5.4cm, height=4cm]{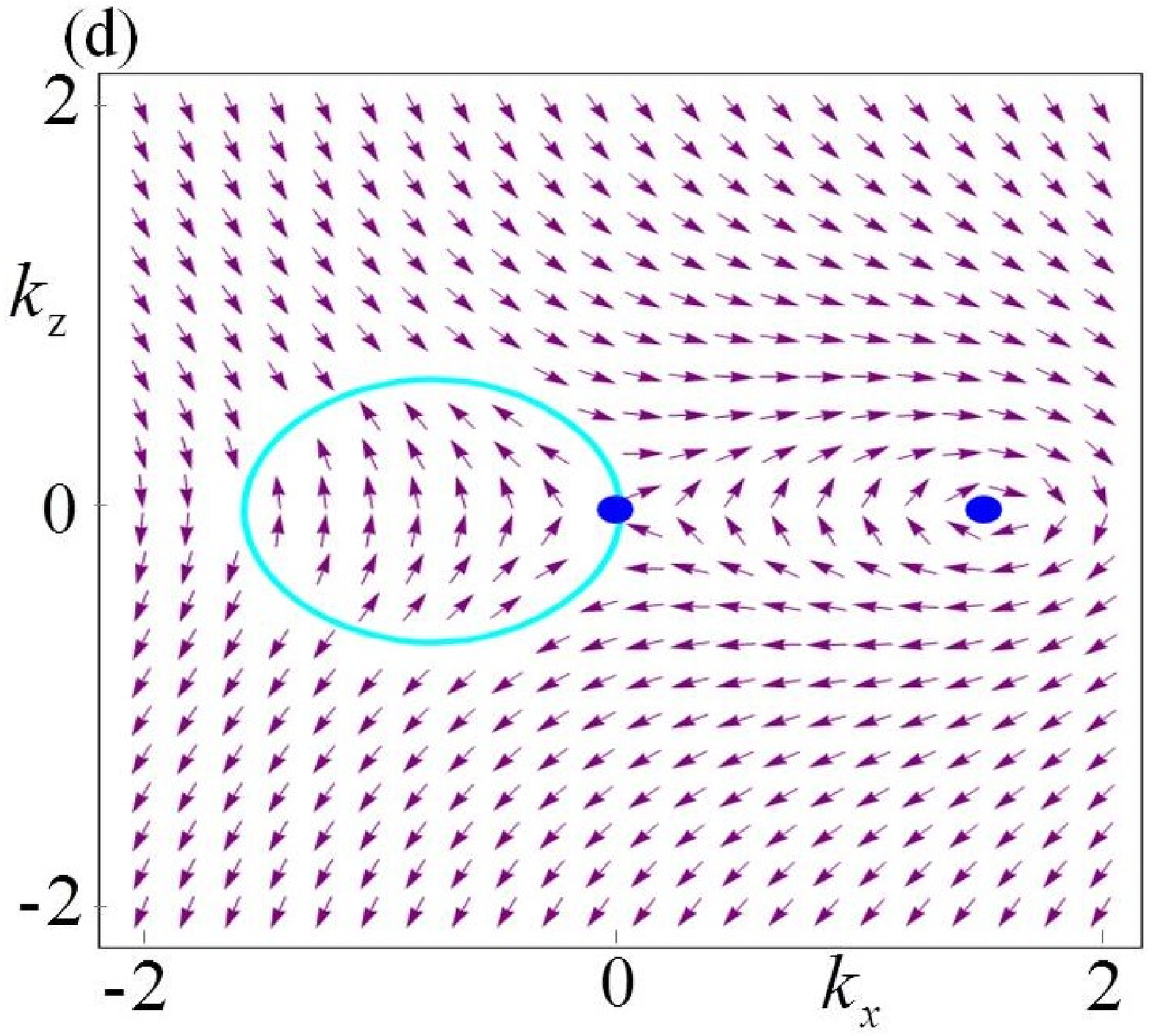}}
\subfigure{\includegraphics[width=5.4cm, height=4cm]{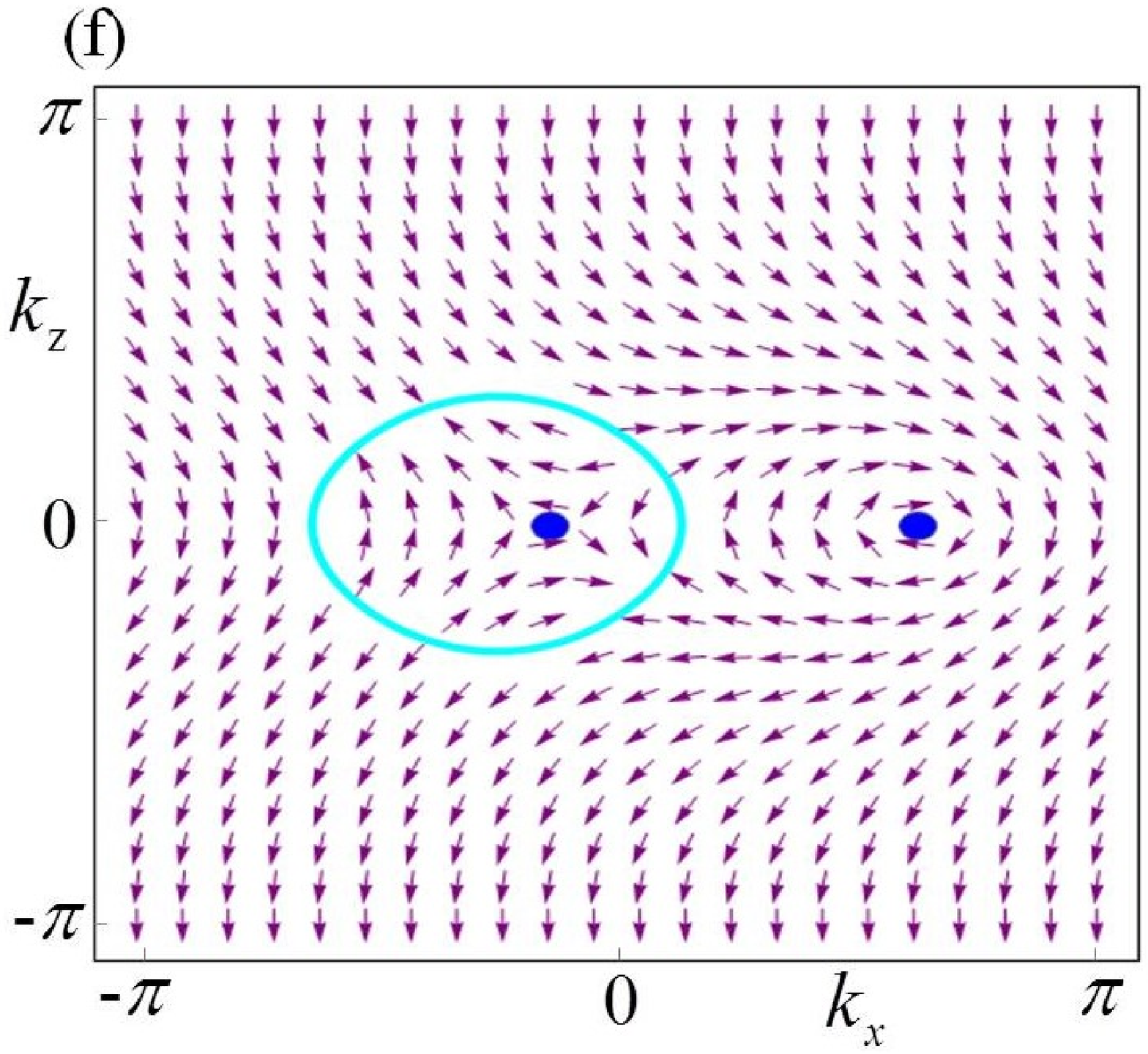}}
\caption{ Nodal rings of Eq.(\ref{xz}), and the pseudospin textures in the $k_y=k_z$ plane, for (a,b) $m_0=3.2$, (c,d) $m_0=3.0$, and (e,f) $m_0=2.5$. The light blue ring locates in the $k_y=k_z$ plane, while the dark blue ring is perpendicular to it. }
\label{rings}
\end{figure*}

Unlike nodal points, nodal rings allow richer topological structures. They can touch at special points\cite{Yu2015,Kim2015,Du2016CaTe,Kobayashi2017}, enabling formations of nodal chains\cite{Bzdusek2016,Yu2017chain}.  In this paper, we introduce new types of topological semimetals, dubbed ``nodal-link semimetals'', which host nontrivially linked nodal rings[e.g., Fig.\ref{rings}(e)]. Furthermore, a method is introduced for constructing two-band models of nodal-link semimetals. We investigate generic physical consequences of nontrivial linking; in particular, a global toroidal $\pi$ Berry phase generates a half-integer shift of Landau level index when the magnetic field is perpendicular to the ring plane. In addition, a suitable periodic external field can drive nodal-link semimetal to a Floquet Hopf insulator.

{\it Models.--}Nodal rings come from the crossing of two adjacent bands, thus we focus on two-band Bloch Hamiltonians, which can generally be written as \bea H(\bk)= a_0(\bk){\bf 1}+ a_1(\bk)\tau_x +a_2(\bk)\tau_y + a_3(\bk)\tau_z, \label{a} \eea where $\bk=(k_x,k_y,k_z)$, $\tau_i$'s are Pauli matrices, and $a_0(\bk)=0$ will be adopted for simplicity (nonzero $a_0$ can be trivially included, if needed). Nodal rings are protected by crystal symmetries. For concreteness, we take the $PT$ symmetry\cite{Fang2015nodal,Zhao2017PT} that ensures the reality of $H(\bk)$, i.e., $a_2(\bk)=0$.  Now the spectra are $E_\pm(\bk)=\pm\sqrt{a_1^2+a_3^2}$, and the nodal rings are given by solving $a_1(\bk)=a_3(\bk)=0$. A purpose of this paper is to construct models with mutually linked nodal rings.

Instead of taking trial-and-error approaches, we put forward a general method based on Hopf maps\cite{wilczek1983,nakahara2003}. They play special roles in quantum spin systems\cite{wilczek1983,fradkin2013}, topological Hopf insulators\cite{moore2008topological,deng2013hopf,deng2016probe,deng2015systematic, kennedy2016,liu2016symmetry}, liquid-crystal solitons\cite{Ackerman2017}, quench dynamics of Chern insulators\cite{wang2016measuring}, and
minimal models for topologically trivial superconductor-based Majorana zero modes\cite{MZM-Hopf}. Mathematically, a Hopf map is a nontrivial mapping from a 3-sphere $S^3$ to a 2-sphere $S^2$, which possesses a nonzero Hopf invariant\cite{wilczek1983,nakahara2003,moore2008topological}. Moreover, it has the geometrical property that the preimage circles of any two points on $S^2$ are linked. Mappings from a 3D torus $T^3$ to $S^2$ inherit the nontrivial topology of Hopf maps through $T^3\rw S^3\rw S^2$, where $T^3\rw S^3$ is a map with unit winding number, and $S^3\rw S^2$ is a Hopf map.

Given any vector function $\bd(\bk)=(d_x,d_y,d_z)$ on the Brillouin zone $T^3$, one can define a mapping from $T^3$ to $S^2$ by $\bk\rw\hat{\bd}(\bk)$, where $\hat{\bd}\equiv\bd/|\bd|$. To define the Hopf invariant,
it is convenient to express the vector $\bd$ in terms of a spinor, namely, $d_{i}(\bk)=z^{\dag}\tau_{i}z$, with $z(\bk)=(z_1,z_2)^T$.   Let us write $z_1=N_1+iN_2, z_2=N_3+iN_4$, then the Hopf invariant simplifies to\cite{liu2016symmetry} \bea n_h=\frac{1}{2\pi^2}\int d^3k \epsilon^{abcd}\hat{N}_a\partial_{k_x}\hat{N}_b \partial_{k_y}\hat{N}_c \partial_{k_z}\hat{N}_d, \eea
where $\hat{N}_a$ is the $a$-th component of the vector ${\bf N}=(N_1,N_2,N_3,N_4)$ normalized to unit length.

Let us take a point on $S^2$, say $\hat{\bn}_1=(0,1,0)$. Under the mapping $\bk\rw\hat{\bd}(\bk)$, all the preimages of $\hat{\bn}_1$ have to satisfy \bea d_x(\bk)=d_z(\bk)=0, \label{preimage} \eea however, the converse is not true, namely, a solution of Eq.(\ref{preimage}) is not necessarily a preimage of $\hat{\bn}_1$. In fact, the preimages of $\hat{\bn}_2=(0,-1,0)$ also satisfy Eq.(\ref{preimage}). In a single equation, Eq.(\ref{preimage}) gives the preimages of both $\hat{\bn}_1$ and $\hat{\bn}_2$. This is among key observations in our construction. When $n_h$ is nonzero, the preimages circles of any two points (say $\hat{\bn}_1$ and $\hat{\bn}_2$) are linked.  Therefore, we can obtain linked nodal rings by taking \bea a_1(\bk)=d_x(\bk),\,\,  a_3(\bk)= d_z(\bk). \label{a1a3} \eea Recall that the $a_2$ term is absent due to crystal symmetries, as discussed above. The same method also works if we start from a different $\bn_{1,2}$; for instance, taking $\bn_{1,2}=(0,0,\pm 1)$ leads to $a_1=d_x, a_3=d_y$, which yields a nodal link as well.

\begin{figure*}[t!]
\subfigure{\includegraphics[width=5.4cm, height=4.1cm]{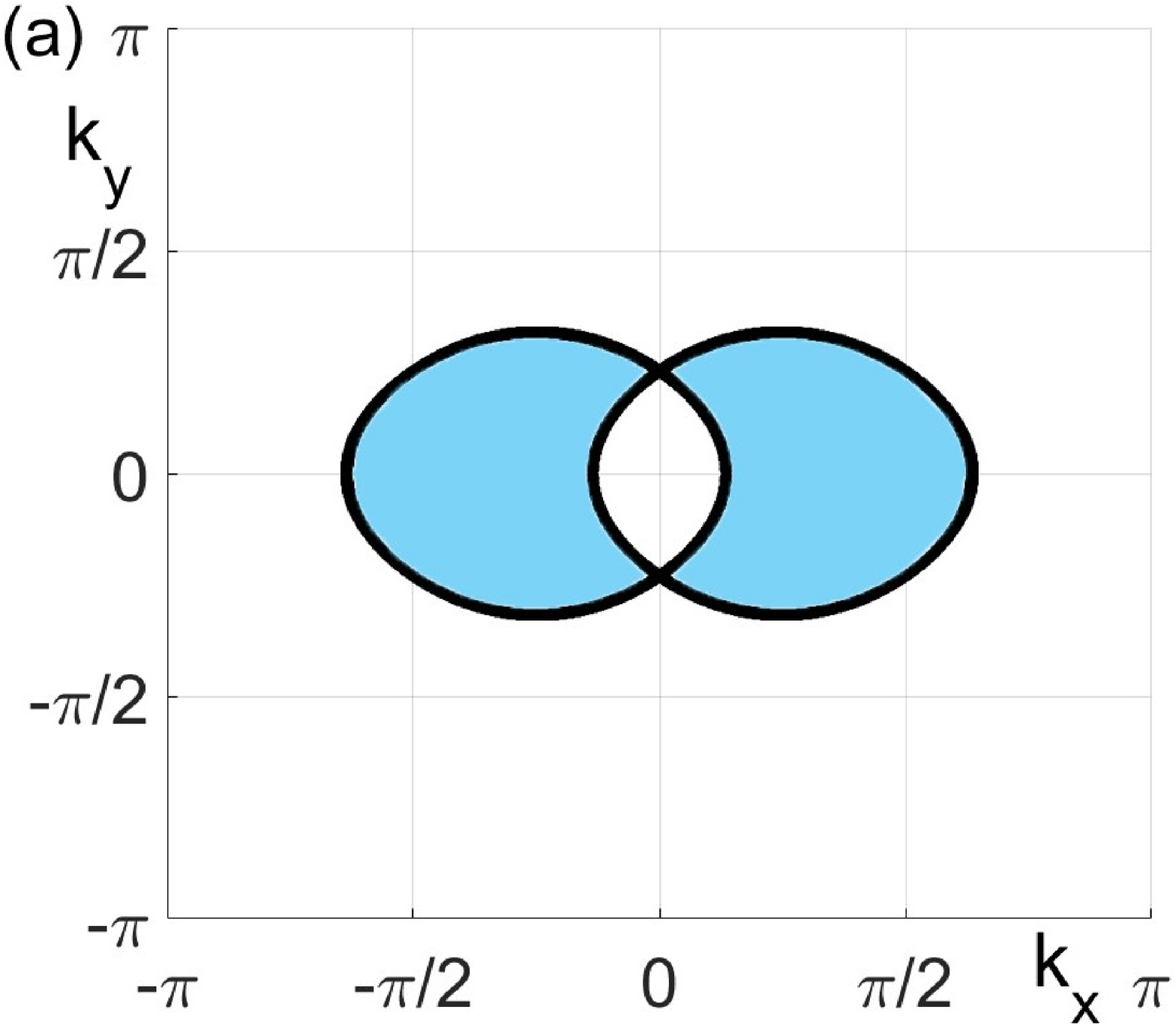}}
\subfigure{\includegraphics[width=5.4cm, height=4.1cm]{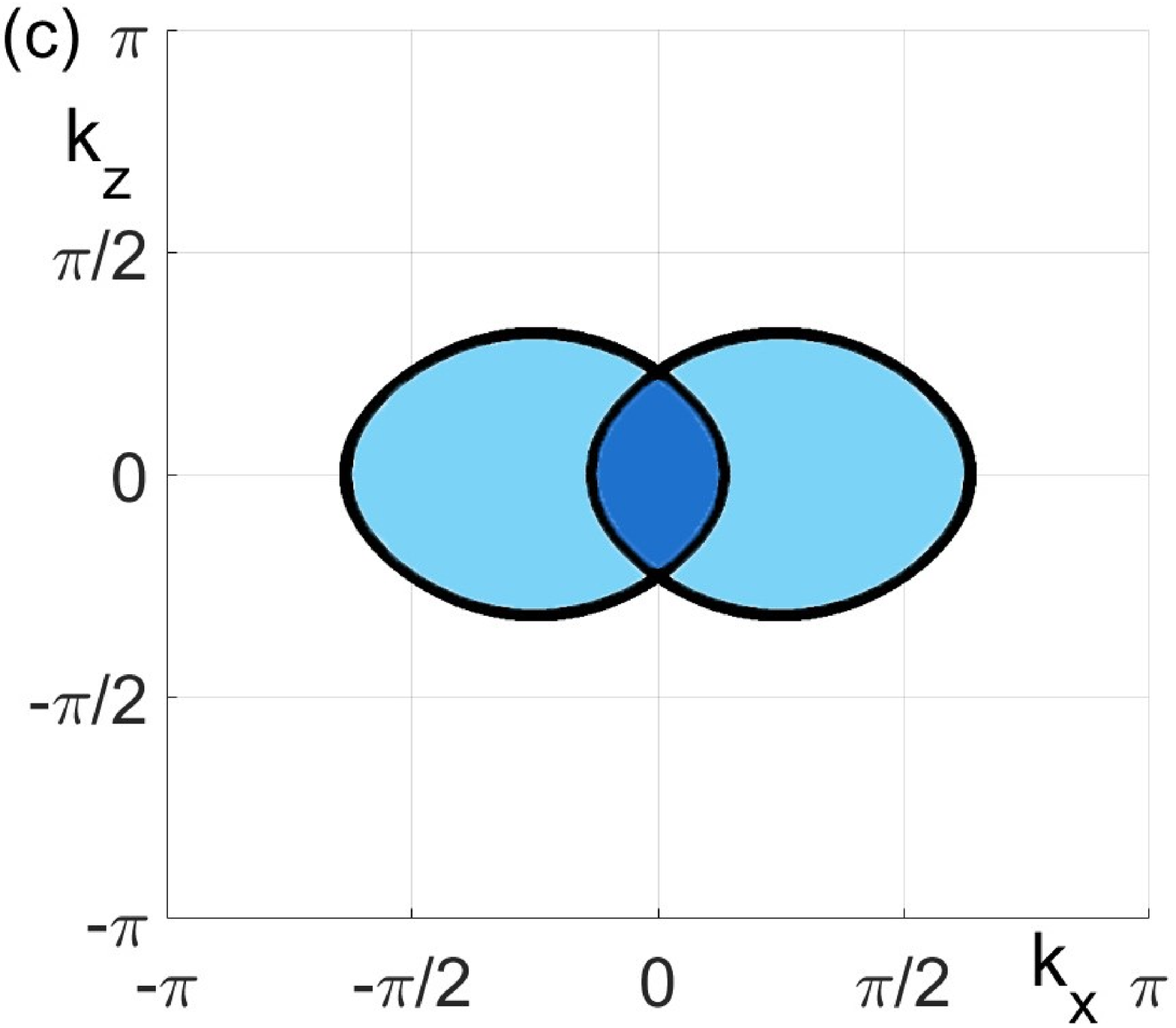}}
\subfigure{\includegraphics[width=5.4cm, height=4.1cm]{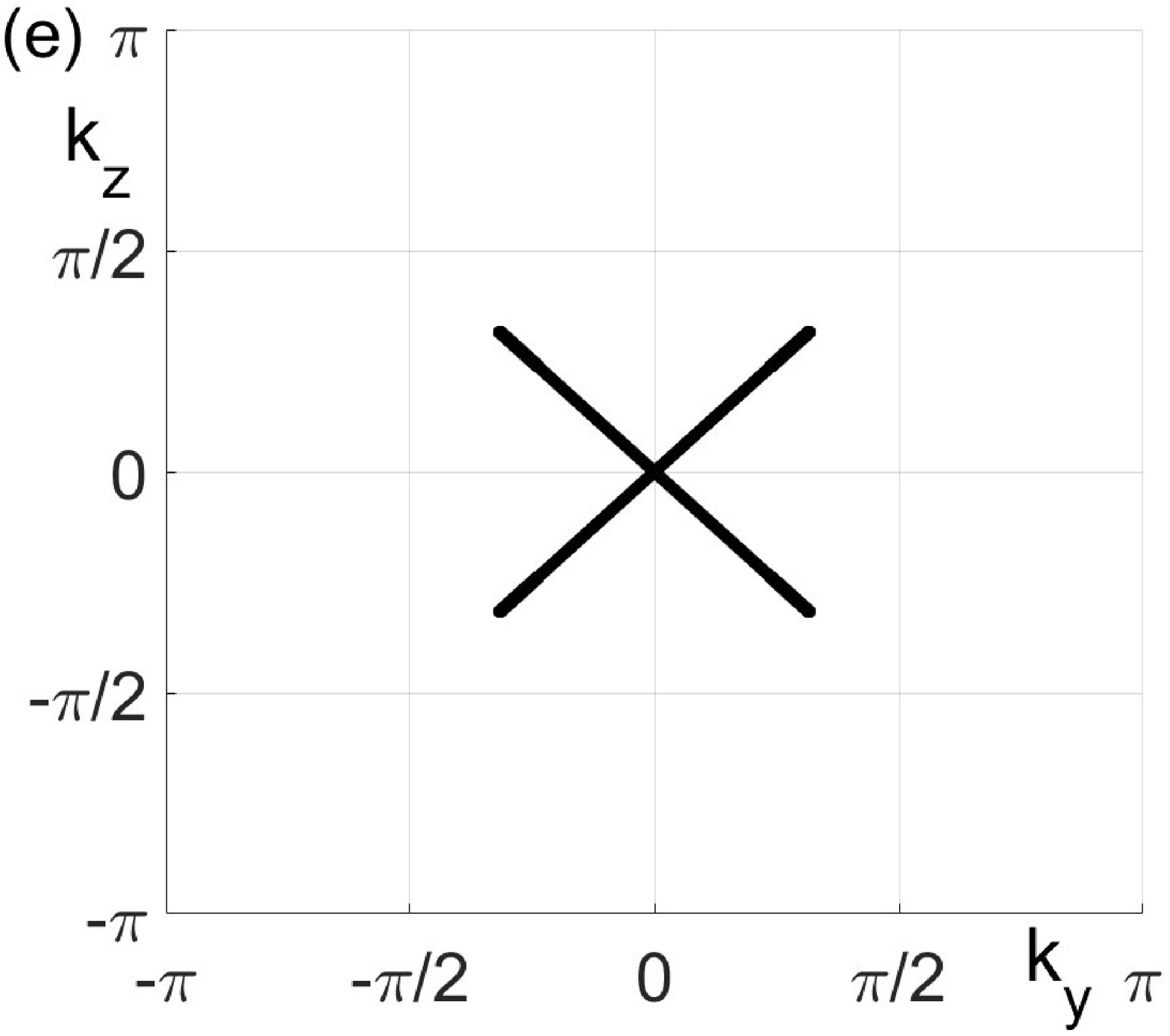}}
\subfigure{\includegraphics[width=5.4cm, height=3.9cm]{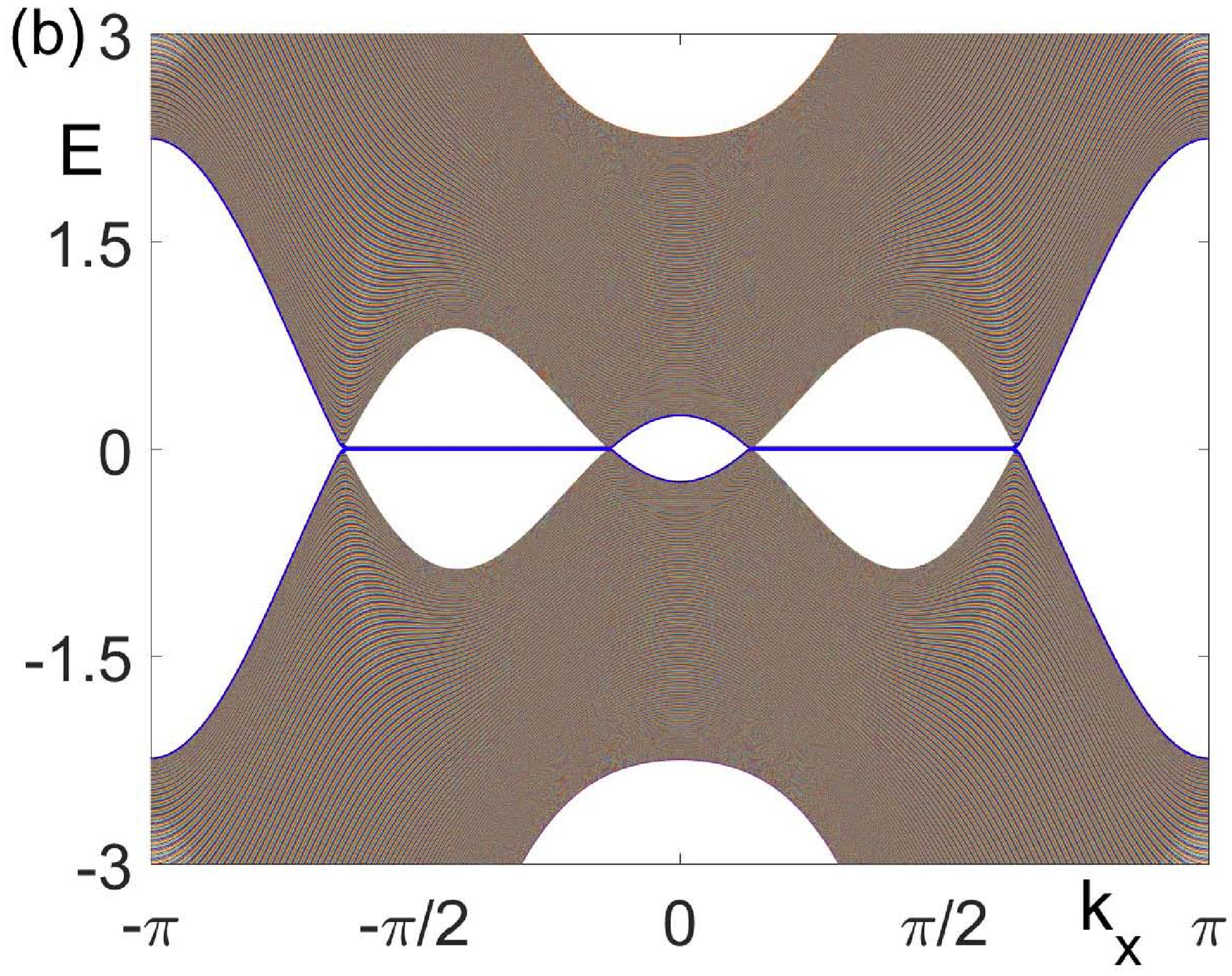}}
\subfigure{\includegraphics[width=5.4cm, height=3.9cm]{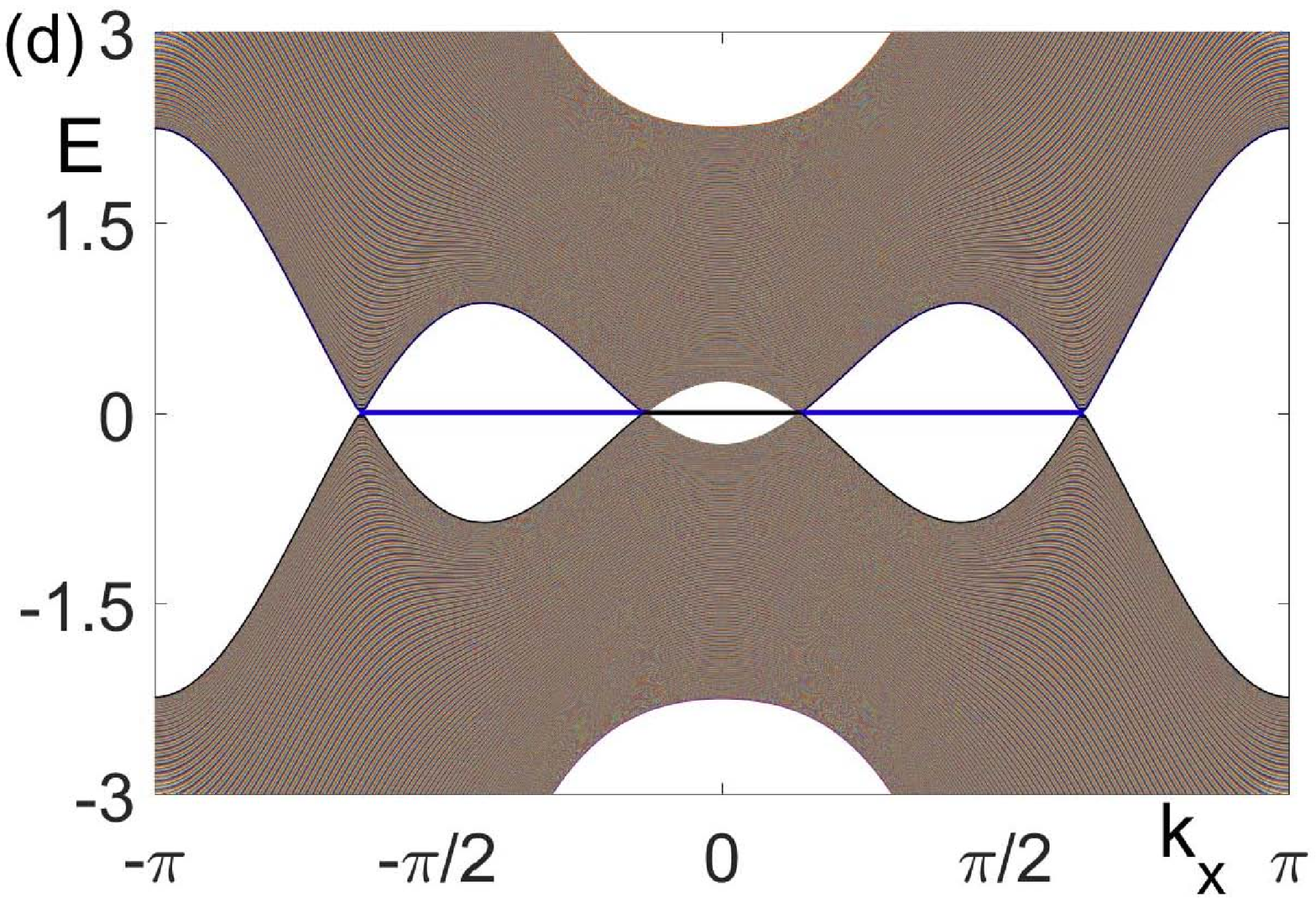}}
\subfigure{\includegraphics[width=5.4cm, height=3.9cm]{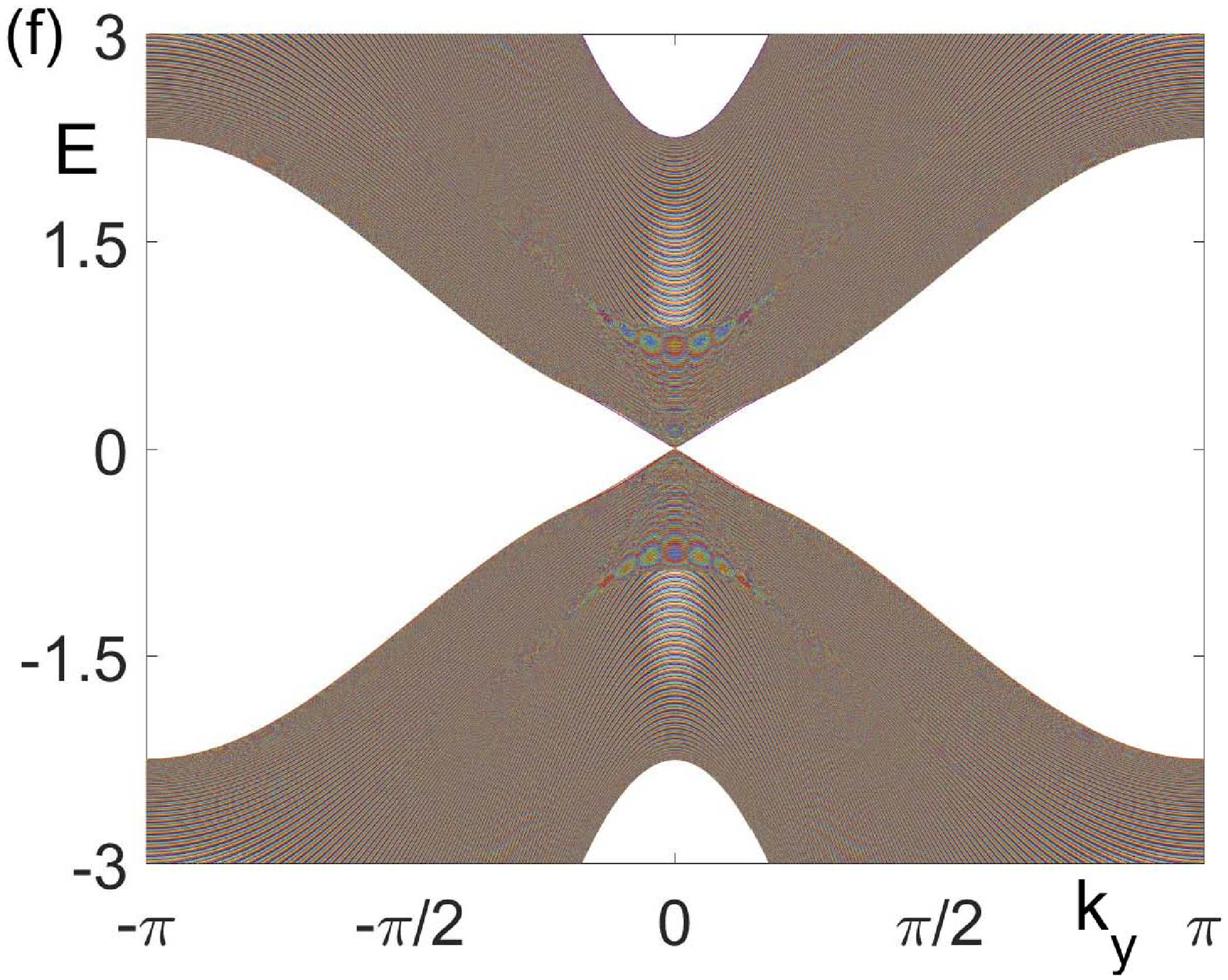}}
\caption{ (a) Surface flat bands for $z=0$ boundary. (b) The spectra as a function of $k_x$ (fixed $k_y=0$) for a 800-site-thick slab perpendicular to $z$ axis.  (c) The $y=0$ surface states. (d) Spectra of a slab perpendicular to $y$ axis. (e) $x=0$ surface Brillouin zone without surface states. (f) Spectra of a slab perpendicular to $x$ axis. The black rings or lines in (a,c,e) are projections of the bulk nodal link to the surface Brillouin zone. In (c), the number of bands in the two-disk-overlapping region is twice that of the non-overlapping regions. Here, $m_0=2.5$. }
\label{surface}
\end{figure*}

This is a quite general approach to construct nodal-link semimetals.
As an example, we take\cite{moore2008topological,deng2013hopf,deng2016probe,deng2015systematic, kennedy2016,liu2016symmetry}
\bea N_1&=&\sin k_x,\, N_2=\sin k_y,\, N_3=\sin k_z,\nn\\ N_4&=&\cos k_x+\cos k_y +\cos k_z -m_0, \eea which has $n_h=-1$ for $1<m_0<3$ and $n_h=0$ for $m_0>3$.
The explicit forms of $d_i$'s read
\begin{eqnarray}
d_{x}&=&2\sin k_{x}\sin k_z+2\sin k_{y}(\sum_{i=x,y,z}\cos k_i -m_{0}),\nonumber\\
d_{y}&=&-2\sin k_{y}\sin k_z +2\sin k_{x}(\sum_{i=x,y,z}\cos k_i -m_{0}),\nonumber\\
d_{z}&=&\sin^{2}k_{x}+\sin^{2}k_{y}-\sin^{2} k_z-(\sum_{i=x,y,z}\cos k_i -m_{0})^{2}. \quad \nn
\end{eqnarray}   Following Eq.(\ref{a1a3}), a lattice model of nodal-link semimetal is \bea H(\bk) &=& [2\sin k_{x}\sin k_z+2\sin k_{y}(\sum_{i=x,y,z}\cos k_i-m_{0})]\tau_x \nn \\ &+& [\sin^{2}k_{x}+\sin^{2}k_{y}-\sin^{2} k_z-(\sum_{i=x,y,z}\cos k_i -m_{0})^{2}]\tau_z. \quad \label{xz} \eea
The nodal rings consist of points where both coefficients of $\tau_x$ and $\tau_z$ vanish. We find that one of the rings is $k_y=k_z, \sin k_x=m_0-\sum_i\cos k_i$, and the other is $k_y=-k_z, \sin k_x=\sum_i\cos k_i-m_0$, shown in light and dark blue, respectively, in Fig.\ref{rings}. Fig.\ref{rings}(a) shows the unlinked rings for $m_0=3.2$, and Fig.\ref{rings}(e) illustrates the linked rings for $m_0=2.5$ (a Hopf link). The critical point $m_0=3.0$, when the two rings cross each other, is shown in Fig.\ref{rings}(c). The direction of pseudospin vector $(d_x,d_z)$ is plotted in Fig.\ref{rings}(b)(d)(f), indicating that the light blue ring encloses a pseudospin vortex in the linked regime [Fig\ref{rings}(f)], in contrast to the unlinked case [Fig\ref{rings}(b)]. The surface states for $m_0=2.5$ are shown in Fig.\ref{surface}. The two-disk-overlapping region has zero and two flat bands in Fig.\ref{surface}(a) and (c), respectively, which is consistent with the winding number\cite{Ryu2002} in each region.

Near the critical point[Fig.\ref{rings}(c)], we can expand the Hamiltonian as: \bea H(\bk)&=&[2k_xk_z+2k_y(m-k^2/2)]\tau_x \nn\\ &+& [k_x^2+k_y^2-k_z^2-(m-k^2/2)^2]\tau_z, \eea where $k^2=\sum_{i=x,y,z}k_i^2$, $m=3-m_0$. At the critical point $m=0$, we have $H(\bk)\approx 2k_xk_z\tau_x +(k_x^2+k_y^2-k_z^2)\tau_z$, thus the dispersion is quadratic in all three directions. Consequently, we find that, for $m=0$, the density of states follows $g(E)\sim\sqrt{E}$ near zero energy, in contrast to $g(E)\sim E$ for $m\neq 0$. Just like that nontrivial topology of insulators can be undone by closing the energy gap, the nontrivial nodal-line linking can be untied through quadratic-dispersion critical points[Fig.\ref{rings}(c)], where $g(E)\sim\sqrt{E}$.

We remark that, although a nodal chain\cite{Bzdusek2016,Yu2017chain} also contains crossings like Fig.\ref{rings}(c), its dispersion is not quadratic in all three directions at the crossing point, and the density of states is linear instead of square-root.

{\it Landau levels.--}A key difference between a usual unlinked ring and a linked one is a global Berry phase along the ring. Let us draw a thin torus enclosing a nodal ring, then the Berry phase along the poloidal direction is always $\pi$; in contrast, the Berry phase along the toroidal direction can be $0$ or $\pi$ (mod $2\pi$), corresponding to the unlinked and linked ring, respectively [The $0$ or $\pi$ toroidal Berry phase of the light-blue ring can be read from the spin texture in Fig.\ref{rings}(b) and (f), respectively].

This $\pi$ toroidal Berry phase can qualitatively affect the Landau levels. It is challenging to find analytic expressions of Landau levels for Eq.(\ref{xz}), nevertheless, a different model allows a simple solution: we take $a_1=d_x(\bk), a_3=d_y(\bk)$ in Eq.(\ref{a}). This Bloch Hamiltonian harbors four
straight nodal lines at $(k_{x},k_{y})=(0,0)$, $(0,\pi)$,
$(0, \pi)$ and $(\pi,\pi)$, respectively. When $1<m_0<3$, a nodal ring encircling $(k_x,k_y)=(0,0)$ is also found as
$\cos k_{x}+\cos k_{y}=m_{0}-1$ in the $k_z=0$ plane. Instead of working on this lattice Hamiltonian, we study its continuum limit for simplicity:
\begin{eqnarray}
H(\bk)=2[k_{x}k_{z}+(m-Ck^{2})k_{y}]\tau_{x}+2[-k_{y}k_{z}+(m-Ck^{2})k_{x}]\tau_{y},\quad \quad \label{xy}
\end{eqnarray}
where $k^{2}=k_{x}^{2}+k_{y}^{2}+k_{z}^{2}$.  $m=3-m_{0}$, and $C=0.5$. We have also done a basis change $\tau_z\rw\tau_y$ for later convenience. This model hosts a nodal line $k_x=k_y=0$ and a nodal ring $k_x^2+k_y^2=m/C$ in the $k_z=0$ plane, which are linked.

Now we add a magnetic field along the $z$ direction, $\bB=B\hat{z}$. It is standard to do the replacement $\bk\rw\mathbf{\Pi}=-i\mathbf{\nabla}+e\bA$
with $\bA=(0,Bx,0)$\footnote{Note that $\Pi_x,\Pi_y$ do not commute. We take the symmetric ordering for operators, e.g., $k_xk_y^2\rw(\Pi_x\Pi_y\Pi_y +\Pi_y\Pi_x\Pi_y +\Pi_y\Pi_y\Pi_x)/3$.}.
It is convenient to introduce the ladder operators,
$a=\frac{l_{B}}{\sqrt{2}}(\Pi_{x}-i\Pi_{y}), \quad a^{\dag}=\frac{l_{B}}{\sqrt{2}}(\Pi_{x}+i\Pi_{y})$,
where $l_{B}=1/\sqrt{eB}$ is the magnetic length. The Hamiltonian becomes
\begin{eqnarray}
H=\left(
    \begin{array}{cc}
      0 & f(k_{z})\frac{\sqrt{2}a^{\dag}}{l_{B}} \\
      \frac{\sqrt{2}a}{l_{B}}f^\dag(k_{z})& 0 \\
    \end{array}
  \right)
\end{eqnarray}
where $f(k_{z})=2[k_{z}-i(m-Ck_{z}^{2}-\frac{2Ca^{\dag}a}{l_{B}^{2}})]$.
The Landau levels are found to be
\begin{eqnarray}
E_{n,\pm}(k_{z})=\left\{\begin{array}{cc}
                   \pm\sqrt{8n[k_{z}^{2}+(m-Ck_{z}^{2}-n\omega_{c})^{2}]}/l_{B}, & n\geq1, \\
                   0, & n=0,
                 \end{array}\right.\label{linkedLL}
\end{eqnarray}
where $\omega_{c}\equiv2C/l_{B}^{2}$. The low energy eigenvalues are around $n\sim 0$ and $n\sim m/\omega_c$, the former coming from the central nodal line at $k_x=k_y=0$, while the latter coming from the nodal ring in the $k_z=0$ plane. As a comparison, we also consider a model with an unlinked ring:
\begin{eqnarray}
H(\bk)=(m-Ck^{2})\tau_{x}+k_{z}\tau_{z}.
\end{eqnarray}
Following the same steps, we find that the Landau levels are
\begin{eqnarray}
E_{n,\pm}(k_{z})=\pm\sqrt{k_{z}^{2}+[m-Ck_{z}^{2}-(n+\frac{1}{2})\omega_{c}]^{2}},n\geq0.\label{normalLL}
\end{eqnarray}
Comparing Eq.(\ref{linkedLL}) and Eq.(\ref{normalLL}), we see that  the presence of a linked nodal line causes a shift of Landau level index by $1/2$, namely, $n\rw n-1/2$, which is a consequence of the $\pi$ toroidal Berry phase. Such a shift can be measured by magneto-transport or magneto-optical experiments.

To highlight the effect of $\pi$ Berry phase, we re-derive the Landau levels using semiclassical quantization\cite{Mikitik1999,onsager1952interpretation,Xiao2010Berry}: \bea S(k_{z})=2\pi eB(n+1/2-\phi_B/2\pi), \eea where $S$ is the cross-sectional area of a $\bk$-space orbit, and $\phi_B$ is the Berry phase along the orbit. For the linked ring, we have $\phi_B=\pi$ (the toroidal Berry phase), and a semiclassical calculation (see Supplemental Material for details) yields the Landau levels in Eq.(\ref{linkedLL}).

{\it Floquet Hopf insulator from nodal link.--}We will show that driving the nodal-link semimetal described by Eq.(\ref{xy}) creates a Floquet Hopf insulator.
We consider a periodic driving generated by a circularly
polarized light(CPL) propagating in $z$ direction. The vector potential $\bA(t)=A_{0}(\cos\omega t, \eta\sin\omega t, 0)$,
where $\eta=1$ and $-1$ stands for right-handed and left-handed
CPL, respectively.
Its effect is described by the minimal
coupling, $H(\bk)\rw H[\bk+e\bA(t)]$. The full Hamiltonian is
time-periodic, $H(\bk, t+T)=H(\bk, t)$ with $T=2\pi/\omega$, thus, it can be expanded as
$H(\bk,t)=\sum_{n}\mathcal{H}_{n}(\bk)e^{in\omega t}$, with
\begin{eqnarray}
\mathcal{H}_{0}(\bk)&=&2[k_{x}k_{z}+k_{y}(\tilde{m}_{2}-Ck^{2})]\tau_{x}\nonumber\\
&&+2[-k_{y}k_{z}+k_{x}(\tilde{m}_{2}-Ck^{2})]\tau_{y},\nonumber\\
\mathcal{H}_{\pm1}(\bk)&=&eA_{0}[k_{z}\mp i\eta(\tilde{m}_{1}-Ck^{2})-2Ck_{y}(k_{x}\mp i\eta k_{y})]\tau_{x}\nonumber\\
&&+eA_{0}[\pm i\eta k_{z}+(\tilde{m}_{1}-Ck^{2})-2Ck_{x}(k_{x}\mp i\eta k_{y})]\tau_{y},\nonumber\\
\mathcal{H}_{\pm2}(\bk)&=&Ce^{2}A_{0}^{2}[(k_{y}\pm i\eta k_{x})\tau_{x}-(k_{x}\mp i\eta k_{y})\tau_{y}],
\end{eqnarray}
where $\tilde{m}_{j=1,2}=m-j Ce^{2}A_{0}^{2}$.
In the off-resonance regime, we can use an effective time-independent Hamiltonian\cite{lindner2011floquet,Kitagawa2011,Oka2009,Inoue2010,
Gu2011,Kitagawa2010a,Kitagawa2010b,
Jiang2011,rudner2013anomalous}:
\begin{eqnarray}
H_{\rm eff}(\bk)&=&\mathcal{H}_{0}+\sum_{n\geq1}\frac{[\mathcal{H}_{n}, \mathcal{H}_{-n}]}{n\omega}+O(\frac{1}{\omega^{2}})\nonumber\\
&=&2[k_{x}k_{z}+k_{y}(\tilde{m}_{2}-Ck^{2})]\tau_{x}+2[-k_{y}k_{z} +k_{x}(\tilde{m}_{2}-Ck^{2})]\tau_{y}\nonumber\\
&&+\lambda[k_{z}^{2}+(\tilde{m}_{1}-Ck^{2})^{2}-2Ck_{\rho}^{2}(\tilde{m}_{1} -Ck^{2}-\frac{\gamma}{4})]\tau_{z}, \label{floquet}
\end{eqnarray}
where $\lambda=4\eta e^{2}A_{0}^{2}/\omega$, $\gamma= Ce^{2}A_{0}^{2}$,
and $k_{\rho}=\sqrt{k_{x}^{2}+k_{y}^{2}}$.
The energy spectrum of $H_{\rm eff}$ is fully gapped.
For weak driving, the minima of the band are located at $k_{z}=0$, $Ck_{\rho}^{2}=\tilde{m}_{2}$,
and the gap is estimated as $E_{g}\approx 12mC(eA_{0})^{4}/\omega$. In the CPL approach, this gap is expected to be rather small. It can be promoted to the order of $(eA_0)^2$ by adding a small $\Delta\tau_z$ pseudo-Zeeman term (see Supplemental Material for details of calculation).

\begin{figure}
\includegraphics[width=8cm, height=6cm]{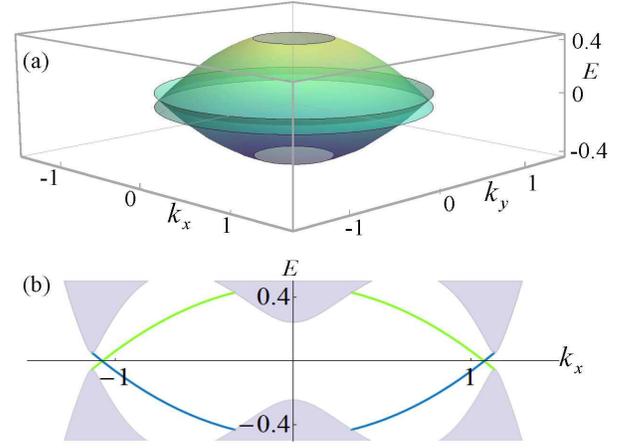}
\caption{ Surface states of a slab perpendicular to the $z$ axis.
Parameters are $\omega=4$, $m=1$, $eA_{0}=0.6$, $C=0.5$. (a) 3D view of surface bands. (b) $E(k_x)$ with
$k_{y}=0$ fixed. The grey regions are bulk bands. }  \label{surfacestate}
\end{figure}

If we remove the $\tau_z$ term, Eq.(\ref{floquet}) hosts a nodal ring and a nodal line linked together. It is readily checked that the coefficient of $\tau_z$ is positive on the line and negative on the ring (for $\eta=+1$), thus, the corresponding unit $\hat{\bd}$ vector is $(0,0,1)$ and $(0,0,-1)$, respectively. This fact suggests that Eq.(\ref{floquet}) describes a (Floquet) Hopf insulators\cite{moore2008topological,deng2013hopf,deng2016probe,deng2015systematic,kennedy2016, liu2016symmetry}, or more precisely, a Floquet Hopf-Chern insulator\cite{kennedy2016}, because the Chern number $C(k_z)=-2$ for arbitrary $k_z$, as found in our numerical calculation. In the definition of Hopf invariant\cite{wilczek1983,moore2008topological}, a nonsingular global Berry potential is needed, which is impossible in the presence of nonzero Chern number, nevertheless, we can study topological surface states. For a slab perpendicular to the $z$ direction, we can solve the differential equation $H_{\rm eff}(k_{x},k_{y},-i\partial_{z})\Psi_{k_{x},k_{y}}(z)=E(k_{x},k_{y})\Psi_{k_{x},k_{y}}(z)$, which gives one surface band for each surface, whose dispersion is (see Supplemental Material for calculation):
$E_{\alpha}(\bk)=\alpha\lambda [\tilde{m}_{2}/C+\gamma^{2}-(1+3C\gamma/2)k_{\rho}^{2}]$,
where $\alpha=+$ $(-)$ for top (bottom) surface. Shown in Fig.\ref{surfacestate}, this dispersion is characteristic of a Hopf insulator\cite{moore2008topological}. For
a fixed $\theta$ (defined as $\theta=\arctan k_{y}/k_{x}$), the surface state of either top or bottom surface is chiral, which can be understood in terms of a Chern number in the $(k_\rho,k_z)$ space\cite{liu2016symmetry}.

{\it Conclusions.--}We have introduced nodal-link semimetals into the family of topological semimetals.  A general method for their model construction has been put forward. These phases may be realized by tuning the hoppings in optical lattices. Finding a solid-state material will be an important progress. It will also be worthwhile to study possible novel effects of linking. Theoretically, our models lay useful ground work for topological field theories\cite{Lian2017Chern} in the Brillouin zone.

{\it Note added.--}After finishing this manuscript, we became aware of a related eprint
by Chen et al\cite{hopflink}, in which a double-helix link is constructed
without using the Hopf map. In another recent work, the
Hopf map method has been generalized by Ezawa to construct
other nodal links\cite{Ezawa2017}.

\emph{Acknowledgements.--}We thank Chen Fang for useful comment. Z.Y., R.B., and Z.W. are supported by NSFC (No. 11674189). Z.Y. is supported in part by China Postdoctoral Science Foundation (2016M590082). L.L is supported by the Ministry of Science and Technology of China (No. 2016YFA0302400) and the National Thousand-Young-Talents Program of China. S.C.Z. is supported by NSF (No. DMR-1305677).

\bibliography{dirac}

\vspace{4mm}

\newpage

{\bf Supplemental Material}

%\vspace{12mm}

\section{Hamiltonian in real space}

The Hamiltonians studied in the main article have been written down in the momentum space. In this supplemental material we
provide their explicit expressions in real space.

For the Hamiltonian with  $a_1=d_{x}(\bk)$ and $a_3=d_{z}(\bk)$ (abbreviated as $d_x$-$d_z$ Hamiltonian), the tight-banding form in real space is
\begin{widetext}
\begin{eqnarray}
H&=&\sum_{lmn}\{[m_{0}(c^{\dag}_{l,m,n}\tau_{z}c_{l+1,m,n}+c^{\dag}_{l,m,n}\tau_{z}c_{l,m+1,n}+c^{\dag}_{l,m,n}\tau_{z}c_{l,m,n+1})+h.c.]
-\frac{1}{2}(c^{\dag}_{l,m,n}\tau_{z}c_{l+2,m,n}+c^{\dag}_{l,m,n}\tau_{z}c_{l,m+2,n}+h.c.)\nonumber\\
&&-\frac{1}{2}(c^{\dag}_{l,m,n}\tau_{z}c_{l+1,m+1,n}+c^{\dag}_{l,m,n}\tau_{z}c_{l+1,m-1,n}
+c^{\dag}_{l,m,n}\tau_{z}c_{l+1,m,n+1}+c^{\dag}_{l,m,n}\tau_{z}c_{l+1,m,n-1}
+c^{\dag}_{l,m,n}\tau_{z}c_{l,m+1,n+1}+c^{\dag}_{l,m,n}\tau_{z}c_{l,m+1,n-1}+h.c.)\nonumber\\
&&-(m_{0}^{2}+1)c^{\dag}_{l,m,n}\tau_{z}c_{l,m,n}+m_{0}(c^{\dag}_{l,m,n}i\tau_{x}c_{l,m+1,n}+h.c.)-
\frac{1}{2}(c^{\dag}_{l,m,n}i\tau_{x}c_{l,m+2,n}+h.c.)
+\frac{1}{2}(c^{\dag}_{l,m,n}\tau_{x}c_{l+1,m,n-1}-c^{\dag}_{l,m,n}\tau_{x}c_{l+1,m,n+1}+h.c.)\nonumber\\
&&+\frac{1}{2}(c^{\dag}_{l,m,n}i\tau_{x}c_{l+1,m-1,n}-c^{\dag}_{l,m,n}i\tau_{x}c_{l+1,m+1,n}+
c^{\dag}_{l,m,n}i\tau_{x}c_{l,m-1,n+1}-c^{\dag}_{l,m,n}i\tau_{x}c_{l,m+1,n+1}+h.c.)\},
\end{eqnarray}
\end{widetext}
where $\{l, m, n\}$ labels the lattice sites along the $\{x, y, z\}$ axis, respectively;
$c_{l,m,n}=(c_{l,m,n;a},c_{l,m,n;b})^{T}$, where $a$ and $b$ denote two degrees of freedom.

For the $d_{x}$-$d_{y}$ Hamiltonian, the tight-banding form in the real space is
\begin{widetext}
\begin{eqnarray}
H&=&\sum_{lmn}\{m_{0}(c^{\dag}_{l,m,n}i\tau_{x}c_{l,m+1,n}+c^{\dag}_{l,m,n}i\tau_{y}c_{l+1,m,n}+h.c.)-
\frac{1}{2}(c^{\dag}_{l,m,n}i\tau_{x}c_{l,m+2,n}+c^{\dag}_{l,m,n}i\tau_{y}c_{l+2,m,n}+h.c.)\nonumber\\
&&+\frac{1}{2}(c^{\dag}_{l,m,n}\tau_{x}c_{l+1,m,n-1}-c^{\dag}_{l,m,n}\tau_{x}c_{l+1,m,n+1}+
c^{\dag}_{l,m,n}\tau_{y}c_{l,m+1,n+1}-c^{\dag}_{l,m,n}\tau_{y}c_{l,m+1,n-1}+h.c.)\nonumber\\
&&+\frac{1}{2}(c^{\dag}_{l,m,n}i\tau_{x}c_{l+1,m-1,n}-c^{\dag}_{l,m,n}i\tau_{x}c_{l+1,m+1,n}+
c^{\dag}_{l,m,n}i\tau_{x}c_{l,m-1,n+1}-c^{\dag}_{l,m,n}i\tau_{x}c_{l,m+1,n+1}+h.c.)\nonumber\\
&&-\frac{1}{2}(c^{\dag}_{l,m,n}i\tau_{y}c_{l+1,m+1,n}+c^{\dag}_{l,m,n}i\tau_{y}c_{l+1,m-1,n}+
c^{\dag}_{l,m,n}i\tau_{y}c_{l+1,m,n+1}+c^{\dag}_{l,m,n}i\tau_{y}c_{l+1,m,n-1}+h.c.)\}.
\end{eqnarray}
\end{widetext}

\section{Semiclassical derivation of Landau levels in a magnetic field}

The semiclassical quantization condition in a magnetic field along the $z$ direction is\cite{Mikitik1999,onsager1952interpretation,Xiao2010Berry} ($\hbar=c=1$)
\begin{eqnarray}
S(k_{z})=2\pi eB(n+\gamma),\label{semiquantization}
\end{eqnarray}
where $S$ denotes the cross-sectional area of the orbital in momentum space,
$n$ is a positive integer, and
$\gamma$ is a constant given by
\begin{eqnarray}
\gamma=\frac{1}{2}-\frac{\phi_B}{2\pi}.
\end{eqnarray}   Here, $\phi_B$ is the Berry phase:
\bea \phi_B=\oint_{\partial S}{\bf a}\cdot d\bk, \eea
where $\partial S$ denotes the boundary of $S$, and ${\bf a}$ is the Berry
connection of the valence band.

The energy spectra of the $d_x$-$d_y$ model $H(\bk)=2[k_{x}k_{z} +(m-Ck^{2})k_{y}]\tau_{x} +2[-k_{y}k_{z}+(m-Ck^{2})k_{x}]\tau_{y}$ are
\begin{eqnarray}
E_{\pm}(\bk)=\pm2\sqrt{(k_{x}^{2}+k_{y}^{2})[k_{z}^{2}+(m-Ck^{2})^{2}]}.
\end{eqnarray}
Because of the rotational symmetry, the $\bk$-space orbital in the presence of magnetic field
along the $z$ direction is a circle in the $k_{x}$-$k_{y}$ plane, in other words, the $\bk$-space
cross-sectional area $S$ of the semiclassical motion is given by \bea S=\pi (k_{x}^{2}+k_{y}^{2}). \eea
As a result,  the energy spectra can be rewritten as
\begin{eqnarray}
E_{\pm}(\bk)=\pm2\sqrt{\frac{S}{\pi}[k_{z}^{2}+(m-Ck_{z}^{2}-C\frac{S}{\pi})^{2}]}. \label{E-S}
\end{eqnarray} For the $d_x$-$d_y$ Hamiltonian in consideration, there is a toroidal Berry phase $\phi_B=\pi$, therefore, $\gamma=0$, and we have, according to Eq.(\ref{semiquantization}): \bea S=2\pi eBn. \eea  Inserting this relation into Eq.(\ref{E-S}), we immediately obtain the
Landau levels in terms of $B$:
\begin{eqnarray}
E_{n,\pm}(k_{z})=\pm\sqrt{8neB[k_{z}^{2}+(m-Ck_{z}^{2}-2nCeB)^{2}]},\, n\geq1, \label{with-pi}
\end{eqnarray} which is consistent with the exact quantum mechanical solutions.
%For $n=0$, we find the Landau level is a flat band without degeneracy.

For the single-nodal-ring model $H(\bk)=(m-Ck^{2})\tau_{x}+k_{z}\tau_{z}$, the toroidal Berry phase $\phi_B=0$, thus we have
$\gamma=1/2$, and \bea S=2\pi eB(n +1/2) \eea according to Eq.(\ref{semiquantization}). The energy spectra of the Hamiltonian are
\begin{eqnarray}
E_{\pm}(\bk)&=&\pm\sqrt{k_{z}^{2}+(m-Ck^{2})^{2}} \nn\\
&=&\pm\sqrt{k_{z}^{2}+(m-Ck_{z}^{2}-C\frac{S}{\pi})^{2}},
\end{eqnarray} where $S=\pi(k_x^2+k_y^2)$ has been used.
With the formula of $S$ inserted, the Landau levels take the form of
\begin{eqnarray}
E_{n,\pm}(k_{z})=\pm\sqrt{k_{z}^{2}+[m-Ck_{z}^{2}-2(n+\frac{1}{2})CeB]^{2}}, \label{without-pi}
\end{eqnarray} which is again consistent with the exact quantum-mechanical solutions.  Notice the difference between $n$ and $n+1/2$ in the Eq.(\ref{with-pi}) and Eq.(\ref{without-pi}). This semiclassical derivation shows that this difference comes from the toroidal Berry phase, which is $\pi$ for a linked ring and $0$ for an unlinked ring.

\section{Surface states of Floquet Hopf insulators}

In the main article, we have obtained an effective time-independent Hamiltonian in the off-resonance regime:

\begin{eqnarray}
H_{\rm eff}(\bk)&=&\mathcal{H}_{0}+\sum_{n\geq1}\frac{[\mathcal{H}_{n},\mathcal{H}_{-n}]}{n\omega}+O(\frac{1}{\omega^{2}})\nonumber\\
&=&2[k_{x}k_{z}+k_{y}(\tilde{m}_{2}-Ck^{2})]\tau_{x}+2[-k_{y}k_{z}+k_{x}(\tilde{m}_{2}-Ck^{2})]\tau_{y}\nonumber\\
&&+\lambda[k_{z}^{2}+(\tilde{m}_{1}-Ck^{2})^{2}-2Ck_{\rho}^{2}(\tilde{m}_{1} -Ck^{2}-\frac{\gamma}{4})]\tau_{z},
\end{eqnarray}
where $\lambda\equiv 4\eta e^{2}A_{0}^{2}/\omega$, $\gamma\equiv Ce^{2}A_{0}^{2}$
and $k_{\rho}=\sqrt{k_{x}^{2}+k_{y}^{2}}$.
It is readily found that the energy spectrum of $H_{\rm eff}$ is fully gapped.

To see how the surface state evolves under the periodic driving,
we treat the photo-induced $\tau_{z}$ term as a perturbation
and consider that the system occupies the $z>0$ region.
Without the $\tau_{z}$ term, the energy
dispersion of the surface state can be obtained by solving the eigenvalue equation
$H_{\rm eff}(k_{x},k_{y},-i\partial_{z})\Psi_{k_{x},k_{y}}(z)=E(k_{x},k_{y})\Psi_{k_{x},k_{y}}(z)$
under the boundary condition $\Psi_{k_{x},k_{y}}(0)=\Psi_{k_{x},k_{y}}(+\infty)=0$.
A straightforward calculation gives
\begin{eqnarray}
\Psi_{k_{x},k_{y}}(z)=\mathcal{N}e^{ik_{x}x+ik_{y}y}e^{-\kappa_{1} z}\sin(\kappa_{2}z)\left(
                                                                   \begin{array}{c}
                                                                     0 \\
                                                                     1 \\
                                                                   \end{array}
                                                                 \right),
\end{eqnarray}
where $\kappa_{1}=1/2C$, $\kappa_{2}=\sqrt{4C(\tilde{m}_{2}-Ck_{\rho}^{2})-1}/2C$,
and $\mathcal{N}$ is an normalization factor. The normalizability requires $\tilde{m}_{2}>Ck_{\rho}^{2}$,
which determines the region within which surface states exist. Without the $\tau_z$ term,
we find that the surface state dispersion is $E(k_{x},k_{y})=0$,
in other words, the energy spectrum of the surface state
is completely flat without the periodic driving. Now we add the photo-induced $\tau_{z}$ term. Based on the first-order perturbation theory,
we find
\begin{eqnarray}
\Delta E(k_{x},k_{y})&=&\int_{0}^{+\infty} dz\Psi_{k_{x},k_{y}}^{*}(z)d_{\rm eff,z}(k_{x},k_{y},-i\partial_{z})\tau_{z}\Psi_{k_{x},k_{y}}(z)\nonumber\\
&=&-\lambda[\tilde{m}_{2}/C +\gamma^{2}-(1+3C\gamma/2)k_{\rho}^{2}]. \label{ssdispersion}
\end{eqnarray}
If the system occupies the $z<0$ region instead of the $z>0$ region, similar procedures lead to
similar dispersion as
Eq.(\ref{ssdispersion}), except that the sign is reversed. Thus, for a finite but sufficiently thick sample, the energy
dispersion for the surface states localized at the surfaces is
\begin{eqnarray}
E_{\alpha}(\bk) = \alpha\lambda [\tilde{m}_{2}/C+\gamma^{2}-(1+3C\gamma/2)k_{\rho}^{2}],
\end{eqnarray}
where $\alpha=+$ $(-)$ for top (bottom) surface. As shown in Fig.\ref{surfacestate},
the two energy bands of the surface state will cross at certain $k_{\rho}$, forming
a nodal ring in the two dimensional momentum space, which is characteristic of a
Hopf insulator.

\begin{figure}
\includegraphics[width=8cm, height=6cm]{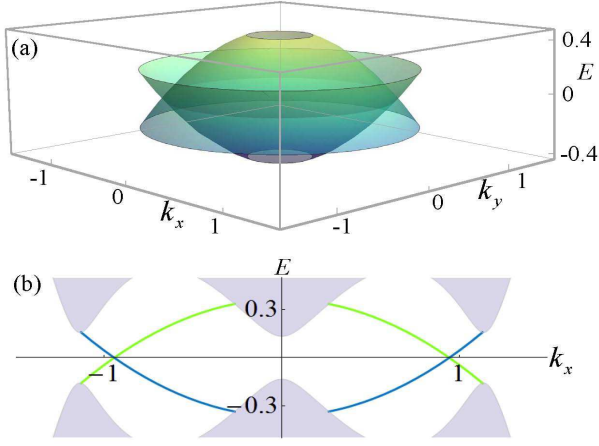}
\caption{ Energy dispersion of the surface states of a slab of Floquet Hopf insulator (with a pseudo-magnetic field $\Delta\tau_z$ added).
Parameters are $\omega=4$, $m=1$, $eA_{0}=0.6$, $C=0.5$, $\Delta=-0.3\lambda$ with $\lambda=4(eA_{0})^{2}/\omega$.
(a) 3D view. (b) $E(k_x)$ with $k_{y}=0$ fixed. The grey regions represent the bulk bands.}  \label{pss}
\end{figure}

As briefly mentioned in the main article, the size of the gap is of the order of $(eA_{0})^{4}$, which is
generally quite small under realistic experimental conditions. This shortcoming can be overcome by introducing
a pseudo-magnetic field $\Delta\tau_{z}$ into $H_{\rm eff}$ (At this stage, this $\Delta\tau_{z}$ term is added phenomenologically. Its physical realization depends on specific system in consideration). As long as $|\Delta|<|\lambda|{\tilde{m}_{1}}^{2}$
and $\eta \Delta<0$, the driven Hamiltonian remains in the nontrivial Hopf insulating phase, but with
the gap modified to $\tilde{E}_{g}=\text{min}\{2\Delta+E_{g}, 2(|\lambda|{\tilde{m}_{1}}^{2}-\Delta)\}$, which is of the order of $(eA_0)^2$.
Moreover, the energy dispersion of the surface states is modified to
\begin{eqnarray}
\tilde{E}_{\alpha}(\bk)=\alpha\left(\lambda [\tilde{m}_{2}/C+\gamma^{2}-(1+3C\gamma/2)k_{\rho}^{2}]+\Delta\right),
\end{eqnarray}
which is shown in Fig.\ref{pss}.

\section{Floquet Hopf insulator from nodal-link semimetal: the lattice model.}

In the main article, we have studied the effect of periodic driving
to the linked nodal rings based on a continuum model. We have obtained a Floquet Hopf insulator. To further corroborate this conclusion, we include here the results for a lattice model.

\begin{figure}
\subfigure{\includegraphics[width=7cm, height=4.5cm]{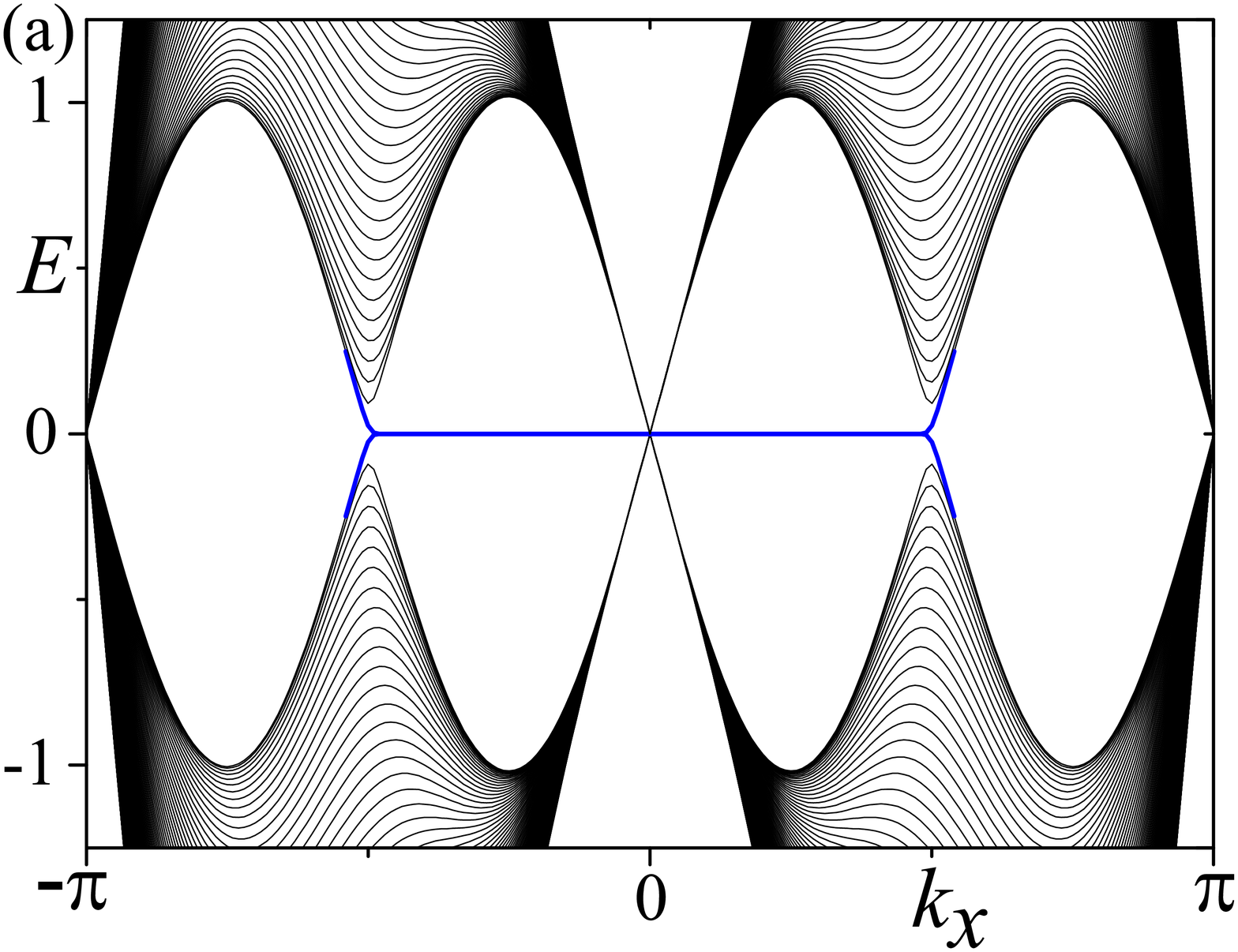}}
\subfigure{\includegraphics[width=7cm, height=4.5cm]{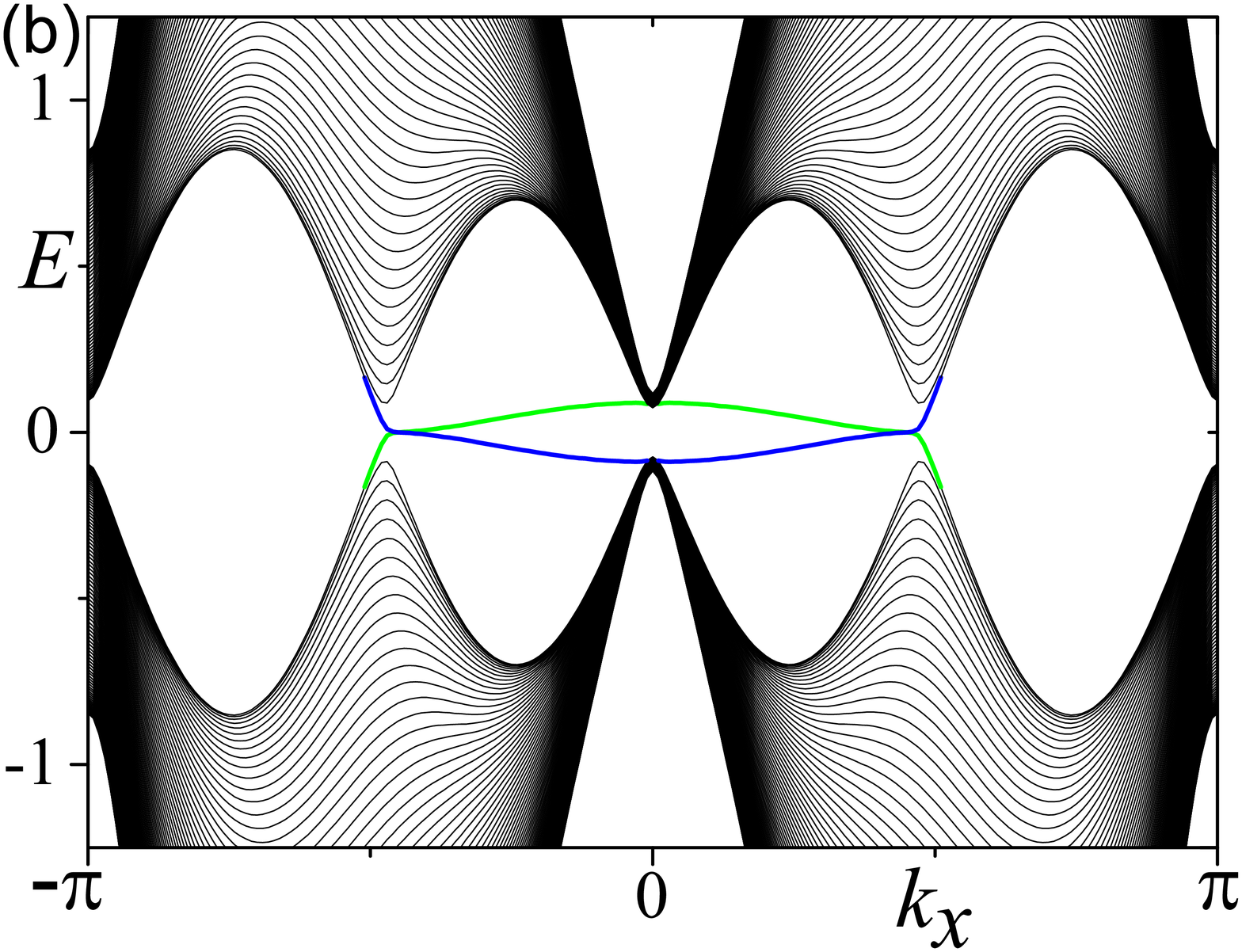}}
\subfigure{\includegraphics[width=7cm, height=4.5cm]{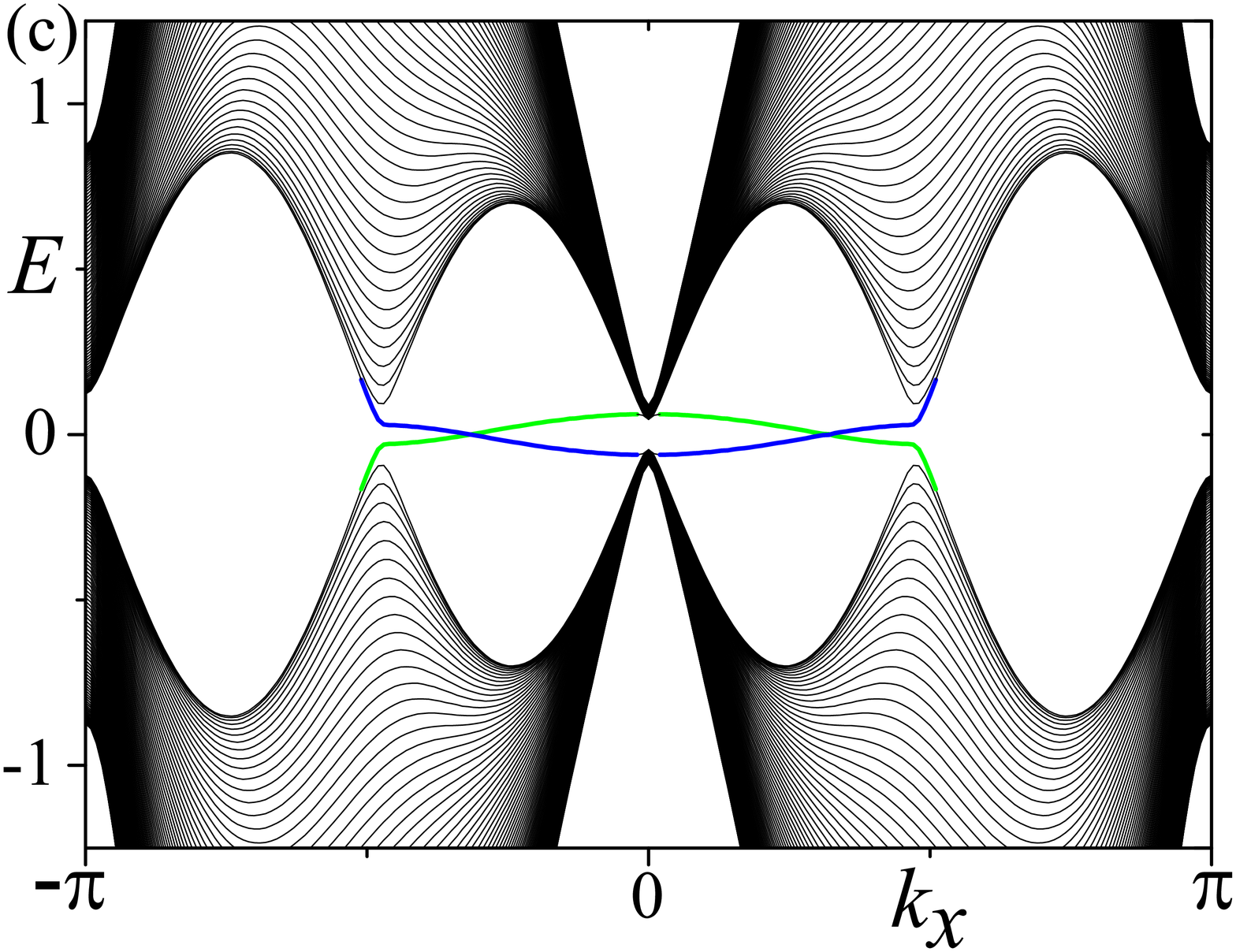}}
\caption{ Surface states evolution of a slab sample under periodic driving. The thickness of slab is 100 site.  $k_y=0$ is fixed. Common parameters: $m_{0}=2$. (a) The static case.
The surface state is of the characteristic ``drumhead'' form.
(b) $eA_{0}=0.5$, $\omega=10$ and $\eta=1$. The green and blue lines denote surface states
on the top and bottom surface, respectively. (c) The parameters are the same as in (b), except that a pseudo-magnetic field term
 $\Delta\tau_{z}$ is added, with $\Delta=-0.3\lambda$. }  \label{latticesurface}
\end{figure}

The lattice Hamiltonian is
\begin{eqnarray}
H(\bk)&=&2[\sin k_{x}\sin k_{z}+\sin k_{y}(\sum_{i}\cos k_{i}-m_{0})]\tau_{x}\nonumber\\
&+& 2[-\sin k_{y}\sin k_{z}+\sin k_{x}(\sum_{i}\cos k_{i}-m_{0})]\tau_{y},
\end{eqnarray} whose continuum limit is studied in the main article.
With the introduction of a circularly
polarized light (CPL) propagating in $z$ direction, the effect of periodic driving is described
by $H(\bk)\rw H[\bk+e\bA(t)]$ with $\bA(t)=A_{0}(\cos\omega t, \eta\sin\omega t, 0)$
with $\eta=\pm 1$ denoting the two kinds of helicity of CPL.
As the full Hamiltonian is time-periodic, $H(\bk, t+T)=H(\bk, t)$ with $T=2\pi/\omega$,  it can be expanded as
$H(\bk,t)=\sum_{n}\mathcal{H}_{n}(\bk)e^{in\omega t}$. By applying the identities for Bessel function,
$\exp[x(\xi-\xi^{-1})/2]=\sum_{n=-\infty}^{\infty}J_{n}(x)\xi^{n}$, $J_{-n}(x)=(-1)^{n}J_{n}(x)$, and
$J_{n}(-x)=(-1)^{n}J_{n}(x)$, we find
\begin{eqnarray}
\sin (k_{x}+eA_{0}\cos\omega t)&=&\sum_{n=-\infty}^{\infty}J_{n}(eA_{0})i^{n}e^{in\omega t}
[-ie^{ik_{x}}+(-1)^{n}ie^{-ik_{x}}]/2,\nonumber\\
\sin (k_{y}+eA_{0}\eta\sin\omega t)&=&\sum_{n=-\infty}^{\infty}J_{n}(eA_{0})e^{in\omega t}\eta^{n}
[-ie^{ik_{y}}+(-1)^{n}ie^{-ik_{y}}]/2,\nonumber\\
\cos (k_{x}+eA_{0}\cos\omega t)&=&\sum_{n=-\infty}^{\infty}J_{n}(eA_{0})i^{n}e^{in\omega t}
[e^{ik_{x}}+(-1)^{n}e^{-ik_{x}}]/2,\nonumber\\
\cos (k_{y}+eA_{0}\eta\sin\omega t)&=&\sum_{n=-\infty}^{\infty}J_{n}(eA_{0})e^{in\omega t}\eta^{n}
[e^{ik_{y}}+(-1)^{n}e^{-ik_{y}}]/2.
\end{eqnarray}
It is straightforward to find that
\begin{eqnarray}
\mathcal{H}_{n}(\bk)&=&\left\{J_{n}(eA_{0})i^{n}[-ie^{ik_{x}}+(-1)^{n}ie^{-ik_{x}}]\sin k_{z}\right.\nonumber\\
&&+J_{n}(eA_{0})\eta^{n}[-ie^{ik_{y}}+(-1)^{n}ie^{-ik_{y}}](\cos k_{z}-m_{0})\nonumber\\
&&+\sum_{l=-\infty}^{\infty}J_{l}(eA_{0})J_{n-l}i^{n-l}\eta^{l}[-ie^{ik_{y}}+i(-1)^{l}e^{-ik_{y}}]\nonumber\\
&&\times[e^{ik_{x}}+(-1)^{n-l}e^{-ik_{x}}]/2+\sum_{l=-\infty}^{\infty}[J_{l}(eA_{0})J_{n-l}(eA_{0})\eta^{n}\nonumber\\
&&\left.\times[-ie^{ik_{y}}+i(-1)^{l}e^{-ik_{y}}][e^{ik_{y}}+(-1)^{n-l}e^{-ik_{y}}]/2\right\}\tau_{x}\nonumber\\
&&+\left\{J_{n}(eA_{0})\eta^{n}[ie^{ik_{y}}-(-1)^{n}ie^{-ik_{y}}]\sin k_{z}\right.\nonumber\\
&&+J_{n}(eA_{0})i^{n}[-ie^{ik_{x}}+(-1)^{n}ie^{-ik_{x}}](\cos k_{z}-m_{0})]\nonumber\\
&&+\sum_{l=-\infty}^{\infty}J_{l}(eA_{0})J_{n-l}(eA_{0})i^{n}[-ie^{ik_{x}}+i(-1)^{l}e^{-ik_{x}}]\nonumber\\
&&\times[e^{ik_{x}}+(-1)^{n-l}e^{-ik_{x}}]/2+\sum_{l=-\infty}^{\infty}J_{l}(eA_{0})J_{n-l}(eA_{0})i^{l}\eta^{n-l}\nonumber\\
&&\times\left.[-ie^{ik_{x}}+i(-1)^{l}e^{-ik_{x}}][e^{ik_{y}}+(-1)^{n-l}e^{-ik_{y}}]/2\right\}\tau_{y}.
\end{eqnarray}
We focus on the regime $eA_{0}<1$ in which $J_{|n|}(eA_{0})$ decrease rapidly with the
increasing of $|n|$, so that it is justified to keep only $\mathcal{H}_{0}$ and $\mathcal{H}_{\pm 1}$,
and ignore terms containing $J_{n}(eA_{0})$ with $|n|\geq2$. The explicit form of $\mathcal{H}_{0}$ and $\mathcal{H}_{\pm 1}$ are
\begin{widetext}
\begin{eqnarray}
\mathcal{H}_{0}(\bk)&=&\left\{2J_{0}(eA_{0})[\sin k_{x}\sin k_{z}+\sin k_{y}(\cos k_{z}-m_{0})]
+2J_{0}^{2}(eA_{0})\sin k_{y}\cos k_{x}\right.\left.+2[J_{0}^{2}(eA_{0})-2J_{1}^{2}(eA_{0})]\sin k_{y}\cos k_{y}\right\}\tau_{x}\nonumber\\
&&+\left\{2J_{0}(eA_{0})[-\sin k_{y}\sin k_{z}+\sin k_{x}(\cos k_{z}-m_{0})]\right.
\left.+2[J_{0}^{2}(eA_{0})-2J_{1}^{2}(eA_{0})]\sin k_{x}\cos k_{x}
+2J_{0}^{2}(eA_{0})\sin k_{x}\cos k_{y}\right\}\tau_{y}\nonumber\\
\mathcal{H}_{+1}(\bk)&=&\left\{2J_{1}(eA_{0})[\cos k_{x}\sin k_{z}-i\eta\cos k_{y}(\cos k_{z}-m_{0})]
-2iJ_{0}(eA_{0})J_{1}(eA_{0})(\eta\cos k_{x}\cos k_{y}-i\sin k_{x}\sin k_{y})\right.\nonumber\\
&&\left.-2i\eta J_{0}(eA_{0})J_{1}(eA_{0})\cos 2k_{y}\right\}\tau_{x}+
\left\{2J_{1}(eA_{0})[i\eta \cos k_{y}\sin k_{z}+\cos k_{x}(\cos k_{z}-m_{0})]\right.\nonumber\\
&&\left.+2J_{0}(eA_{0})J_{1}(eA_{0})\cos 2k_{x}+2J_{0}(eA_{0})J_{1}(eA_{0})(\cos k_{x}\cos k_{y}+i\eta\sin k_{x}\sin k_{y})\right\}\tau_{y},\nonumber\\
\mathcal{H}_{-1}(\bk)&=&\left\{2J_{1}(eA_{0})[\cos k_{x}\sin k_{z}+i\eta\cos k_{y}(\cos k_{z}-m_{0})]
+2iJ_{0}(eA_{0})J_{1}(eA_{0})(\eta \cos k_{x}\cos k_{y}+i\sin k_{x}\sin k_{y})\right.\nonumber\\
&&\left.+2i\eta J_{0}(eA_{0})J_{1}(eA_{0})\cos 2k_{y}\right\}\tau_{x}+
\left\{2J_{1}(eA_{0})[-i\eta \cos k_{y}\sin k_{z}+\cos k_{x}(\cos k_{z}-m_{0})]\right.\nonumber\\
&&\left.+2J_{0}(eA_{0})J_{1}(eA_{0})\cos 2k_{x}+2J_{0}(eA_{0})J_{1}(eA_{0})(\cos k_{x}\cos k_{y}-i\eta \sin k_{x}\sin k_{y})\right\}\tau_{y}.
\end{eqnarray}
\end{widetext}
We consider the off-resonant regime, in which the effective time-independent
Hamiltonian is given by
\begin{widetext}
\begin{eqnarray}
H_{\rm eff}(\bk)&=&\mathcal{H}_{0}+\frac{[\mathcal{H}_{+1},\mathcal{H}_{-1}]}{\omega}+O(\frac{1}{\omega^{2}})\nonumber\\
&=&\left\{2J_{0}(eA_{0})[\sin k_{x}\sin k_{z}+\sin k_{y}(\cos k_{z}-m_{0})]
+2J_{0}^{2}(eA_{0})\sin k_{y}\cos k_{x}\right.\left.+2[J_{0}^{2}(eA_{0})-2J_{1}^{2}(eA_{0})]\sin k_{y}\cos k_{y}\right\}\tau_{x}\nonumber\\
&&+\left\{2J_{0}(eA_{0})[-\sin k_{y}\sin k_{z}+\sin k_{x}(\cos k_{z}-m_{0})]\right.
\left.+2[J_{0}^{2}(eA_{0})-2J_{1}^{2}(eA_{0})]\sin k_{x}\cos k_{x}
+2J_{0}^{2}(eA_{0})\sin k_{x}\cos k_{y}\right\}\tau_{y}\nonumber\\
&&+\lambda\left\{\cos k_{x}\cos k_{y}[(\cos k_{z}-m_{0})^{2}+\sin^{2}k_{z}]+J_{0}(eA_{0})(\cos 2k_{x}\cos k_{y}+\cos k_{x}\cos 2k_{y}
+\cos^{2}k_{x}\cos k_{y}+\cos k_{x}\cos^{2} k_{y})\right.\nonumber\\
&&\times(\cos k_{z}-m_{0})+J_{0}(eA_{0})(\cos k_{x}-\cos k_{y})\sin k_{x}\sin k_{y}\sin k_{z}\nonumber\\
&&\left.+J_{0}^{2}(eA_{0})[\cos2k_{x}\cos2k_{y}+
\cos k_{x}\cos k_{y}(\cos2k_{x}+\cos 2k_{y})+\cos^{2}k_{x}\cos^{2}k_{y}-\sin^{2}k_{x}\sin^{2}k_{y}]\right\}\tau_{z},
\end{eqnarray}
\end{widetext}
where $\lambda=16\eta J_{1}^{2}(eA_{0})/\omega$.

Now we study the surface states. We consider a slab perpendicular to the $z$ direction, with
periodic boundary condition in the $x$ and $y$ directions. The numerical
results are shown in Fig.\ref{latticesurface}. Without periodic driving,
the energy dispersion of the surface state is found to be completely flat,
as shown in Fig.\ref{latticesurface}(a). In the presence of periodic driving,
the surface state becomes dispersive. Besides, the two energy bands of the surface
states (from top and bottom surfaces) are found to cross each other along a closed line, which is a characteristic
property of Hopf insulators.
After adding a pseudo-magnetic field $\Delta\tau_{z}$ to $H_{\rm eff}$ (note that such a constant term
does not affect the form of the photo-induced part of the effective Hamiltonian), the crossing of the
two surface energy bands becomes more prominent, as shown in Fig.\ref{latticesurface}(c).
It is apparent that the above results for the lattice model qualitatively agrees with that of
the continuum model.

\end{document}